\documentclass[a4paper,fleqn,usenatbib]{mnras}

\usepackage[T1]{fontenc}
\usepackage{ae,aecompl}

\usepackage{color,soul}
\usepackage{rotating}
\usepackage{soul}
\usepackage{fancyvrb}

\sethlcolor{green}

\usepackage{lineno}

\newcommand{\HOSTLIB}{{\tt HOSTLIB}}
\newcommand{\SNANA}{{\tt SNANA}}
\newcommand{\SMP}{{\tt SMP}}

\newcommand{\SALTII}{{\sc SALT-II}}
\newcommand{\SDSS}{SDSS-II}

\newcommand{\Diff}{{\tt DiffImg}}

\newcommand{\DESSN}{DES-SN}
\newcommand{\DESLOWZ}{{\sc DES-SN3YR}}  

\newcommand{\lowz}{low-$z$}

\newcommand{\spec}{spectroscopic}
\newcommand{\specy}{spectroscopically}
\newcommand{\obs}{observation}
\newcommand{\obss}{observations}
\newcommand{\eff}{efficiency}

\newcommand{\unc}{uncertainty}
\newcommand{\uncs}{uncertainties}

\newcommand{\MJD}{{\bf\tt MJD}}

\newcommand{\GAIN}{{\bf\tt GAIN}}

\newcommand{\SKYSIG}{{\bf\tt SKYSIG}}
\newcommand{\PSFSIG}{\sigma_{\tt PSF}}
\newcommand{\ZPTADU}{{\bf\tt ZPTADU}}
\newcommand{\ZPTpe}{{\bf\tt ZPTpe}}
\newcommand{\NEA}{{\bf\tt NEA}}

\newcommand{\ZPHOT}{{\bf\tt ZPHOT}}
\newcommand{\ZPHOTERR}{{\bf\tt ZPHOTERR}}

\newcommand{\Sersic}{S\'ersic}

\newcommand{\OL}{\Omega_{\Lambda}}
\newcommand{\OM}{\Omega_{\rm M}}
\newcommand{\zobs}{z_{\rm obs}}
\newcommand{\zcmb}{z_{\rm cmb,true}}
\newcommand{\zcmbObs}{z_{\rm cmb,obs}}
\newcommand{\zhel}{z_{\rm hel,true}}
\newcommand{\zhelObs}{z_{\rm hel,obs}}
\newcommand{\mutrue}{\mu_{\rm true}}

\newcommand{\mubias}{$\mu$-bias}

\newcommand{\M}{{\cal M}}
\newcommand{\mumodel}{\mu_{\rm model}}
\newcommand{\muLens}{\mu_{\rm lens}}
\newcommand{\muLensz}{\mu_{{\rm lens},z} }
\newcommand{\muLensA}{\mu_{{\rm lens},0.4} }
\newcommand{\muLensB}{\mu_{{\rm lens},0.6} }
\newcommand{\vpec}{v_{\rm pec}}
\newcommand{\vcor}{v_{\rm pec,cor}}
\newcommand{\verr}{v_{\rm pec,err}}
\newcommand{\sigmavpec}{\sigma_{\rm vpec}}
\newcommand{\sigz}{\sigma_{z}}
\newcommand{\dzNoise}{\delta z_{\rm noise}}

\newcommand{\mtrue}{m_{\rm true}}
\newcommand{\Ftrue}{F_{\rm true}}
\newcommand{\FSMP}{F_{\rm\tt SMP}}
\newcommand{\FSIM}{F_{\rm sim}}
\newcommand{\sigF}{\sigma_{F}}
\newcommand{\sigFP}{\sigma_{F}^{\prime}}
\newcommand{\sigFtrue}{\sigma_{\rm Ftrue}}
\newcommand{\sigHOST}{\sigma_{\rm host}}

\newcommand{\ERRSCALESIM}{\hat{S}_{\rm sim}}
\newcommand{\ERRSCALESMP}{\hat{S}_{\rm\tt SMP}}
\newcommand{\zSN}{z_{\rm SN}}
\newcommand{\zhelSN}{z_{\rm SN,hel}}
\newcommand{\zhelHOST}{z_{\rm HOST,hel}}

\newcommand{\EFFTOT}{E_{\rm TOT}}  
\newcommand{\vecSNR}{\vec{\rm\bf SNR} }
\newcommand{\EFFSNR}{E_{\vecSNR}}
\newcommand{\EFFspec}{E_{\rm spec}}  
\newcommand{\ipeak}{i_{\rm peak}}
\newcommand{\Bpeak}{B_{\rm peak}}
\newcommand{\Trest}{T_{\rm rest}}

\newcommand{\GaussSigF}{{\cal N}(0,\sigF)}

\newcommand{\sigplus}{\sigma_{+}}
\newcommand{\sigminus}{\sigma_{-}}
\newcommand{\cpeak}{\bar{c}}

\newcommand{\SFR}{\rm SFR}
\newcommand{\Ovec}{\vec{\cal O}}
\newcommand{\mSB}{m_{\rm SB}}

\newcommand{\sigSMP}{\sigma_{\rm\tt SMP}}

\newcommand{\NDES}{207}
\newcommand{\NLOWZ}{122}
\newcommand{\NTOT}{329}
\newcommand{\NFAKE}{10,000}
\newcommand{\SIGMAVPEC}{300}
\newcommand{\SIGMAVCOR}{250}
\newcommand{\sigzOZDES}{10^{-4}}

\newcommand{\NSIMBIASCOR}{1.3{\times}10^6}
\newcommand{\MAXMUBIASALL}{0.05}
\newcommand{\MAXMUBIAS}{0.4}

\newcommand{\Mval}{-19.365}
\newcommand{\urlDR}{{\tt https://des.ncsa.illinois.edu/releases/sn}}

\title[Simulations to Correct SN~Ia Distance Biases]{ First Cosmology Results using Type Ia Supernova from the Dark Energy Survey:
        Simulations  to Correct Supernova Distance Biases }
   
\author[DES Collaboration]{
\parbox{\textwidth}{
\Large
R.~Kessler$^{1,2}$,
D.~Brout$^{3}$,
C.~B.~D'Andrea$^{3}$,
T.~M.~Davis$^{4}$,
S.~R.~Hinton$^{4}$,
A.~G.~Kim$^{5}$,
J.~Lasker$^{1,2}$,
C.~Lidman$^{6}$,
E.~Macaulay$^{7}$,
A.~M\"oller$^{8,6}$
% ...
\begin{center} (DES Collaboration) \end{center}
}
\vspace{0.4cm}
\\
\parbox{\textwidth}{
Affiliations are in Appendix~\ref{sec:affil}.
}
}

\date{Accepted XXX. Received YYY}
\pubyear{2015}

\begin{document} 
\label{firstpage}
\pagerange{\pageref{firstpage}--\pageref{lastpage}}
\maketitle


\begin{abstract}
We describe catalog-level simulations of Type Ia supernova (SN~Ia) light curves
in the  Dark Energy Survey Supernova Program (\DESSN), 
and in low-redshift samples from the Center for Astrophysics (CfA)
and the Carnegie Supernova Project (CSP).
These simulations are used to model biases from selection effects
and light curve analysis, and to determine bias corrections for SN~Ia 
distance moduli that are used to measure cosmological parameters.
To generate realistic light curves, the simulation uses a detailed SN~Ia model,
incorporates information from \obss\ (PSF, sky noise, zero point),
and uses summary information  (e.g., detection efficiency vs. signal to noise ratio)
based on \NFAKE\ fake SN light curves whose fluxes 
were overlaid on images and processed with our analysis pipelines.
The quality of the simulation is illustrated by predicting distributions 
observed in the data.
Averaging within redshift bins,
we find distance modulus biases up to $\MAXMUBIASALL$~mag 
over  the redshift ranges of the \lowz\ and \DESSN\ samples.
For individual events, particularly those with extreme red or blue color,
distance biases can reach  $\MAXMUBIAS$~mag.
Therefore, accurately determining bias corrections is critical for precision 
measurements of cosmological parameters.
Files used to make these corrections are available at {\urlDR}.
\end{abstract}

\begin{keywords}
techniques --  cosmology -- supernovae
\end{keywords}

 \section{Introduction}
 \label{sec:intro}
  
Since the discovery of cosmic acceleration \citep{Riess98,Saul99}
using a few dozen Type Ia supernovae (SNe~Ia),  surveys have
been collecting larger SN~Ia samples and improving the 
measurement precision of the dark energy equation of state parameter ($w$).
This improvement is in large part due to the use of rolling surveys to 
discover and measure large numbers of SN~Ia light curves 
in multiple passbands with the same instrument.
The most recent Pantheon sample \citep{Pantheon} includes
more than 1,000 \specy\ confirmed SNe~Ia from  low and high redshift surveys.
Compared to the 20th century sample used to discover cosmic acceleration,
the Pantheon sample has more than a 20-fold increase in statistics
and much higher quality light curves.

In addition to improving statistics and light curve quality, reducing systematic
\uncs\ is equally important. While most of the attention is on calibration,
which is the largest source of systematic \unc, significant effort 
over more than a decade
has gone into making robust simulations that are used to correct for the redshift-dependent 
distance-modulus bias  ({\mubias}) 
arising from  selection effects.
Selection effects include several 
sources of experimental inefficiencies:
instrumental magnitude limits resulting in Malmquist bias, 
detection requirements from an image-subtraction pipeline used to discover transients, 
target selection for \spec\ follow-up, and 
cosmology-analysis requirements.
These selection effects introduce average \mubias\ variations reaching $\sim 0.05$~mag
at the high-redshift range of a survey.
(e.g., see Fig.~5 in \citet{JLA} and Fig.~6 in \citet{Pantheon}),
and the \mubias\  averaged in specific color ranges 
can be an order of magnitude larger.

In addition to  sample selection, the \mubias\ depends on 
the parent populations of the SN~Ia stretch and color, and also on intrinsic brightness variations,
hereafter called `intrinsic scatter,'
in both the absolute magnitude and in the colors.
For precision measurements of cosmological parameters,
simulations are essential to determine \mubias\ corrections,
and these simulations require accurate models of SN light curves and  sample selection.

The main focus of this paper is to describe our simulations of 
\specy\ confirmed  SNe~Ia from three seasons of the
Dark Energy Survey Supernova Program (\DESSN),
and the associated \lowz\ sample. 
The combination of these two samples, called \DESLOWZ, is used to measure cosmological
parameters presented in \cite{KEY2018}. 
All simulations were performed with the public 
``SuperNova ANAlysis'' (\SNANA) software package 
\citep{SNANA}.\footnote{\tt https://snana.uchicago.edu}
In addition to SNe~Ia, 
a variety of source models can be supplied to
the \SNANA\ simulation, including core collapse (CC) SNe, kilonovae (KN),
or any rest-frame model described by a time-dependent sequence of 
spectral energy distributions.

The \SNANA\ simulations are performed at the ``catalog level,'' which means
that rather than simulating SN light curves on images, 
light curve fluxes and \uncs\ are computed from image properties.
The simulation inputs include a rest-frame source model,  volumetric rate versus redshift,
cosmological parameters (e.g., $\OM,w$), telescope transmission in each passband,
calibration reference, observing and image properties from a survey,
and random numbers to generate Poisson fluctuations.
The simulated light curves are treated like calibrated light curves from a survey,
and are thus analyzed with the same software as for the data.

The \SNANA\ simulation is ideally suited for rolling searches in which
the same instrument is used for both discovery and for measuring light curves.
Surveys with rolling searches include 
the Supernova Legacy Survey (SNLS; \citealt{Astier2006}),
the Sloan Digital Sky Survey-II (SDSS-II; \citealt{Frieman2008,Sako2018}),
the Panoramic Survey Telescope and Rapid Response System 
(PS1; \citealt{PS_2002}), and DES.
The \lowz\ sample, however,  is based on follow-up \obss\ from independent
search programs
\citep[CFA, CSP]{CFA3_Hicken2009,CFA4_Hicken2012,CSP_Contreras2010,CSP_Fol2010}, 
and the observing properties of the search are not
available to perform a proper simulation. The \lowz\ simulation, therefore,
requires additional assumptions and approximations.

Simulated corrections first appeared in the SNLS cosmology analysis 
\citep{Astier2006}.
\citet{K09} analyzed several samples ({\lowz}, \SDSS, SNLS, ESSENCE),
which led to a more general  \SNANA\  framework to simulate 
\mubias\ corrections  for  arbitrary surveys.  
The heart of this framework is a set of two libraries. 
First, an \obs\ library where each observation date includes a characterization of the
point spread function (PSF), sky and readout noise, template noise, zero point, and gain.
Second, a host-galaxy library includes magnitudes and surface profiles,
and is used to compute Poisson noise and to model the local surface brightness.
For a specified light curve model, these libraries are used to convert
top-of-the-atmosphere model magnitudes into observed fluxes and \uncs.

After a survey has completed, assembling the libraries is a relatively straightforward exercise, 
and \SNANA\ simulations have been used in numerous cosmology analyses 
\citep{K09,Conley2011,JLA, Rest2014, Scolnic2014cosmo,Pantheon}.
Before a survey has started, predicting the libraries
is one of the critical tasks for making reliable forecasts.
Such pre-survey forecasts with the \SNANA\ simulation have been made for 
LSST\footnote{Large Synoptic Survey Telescope: {\tt https://www.lsst.org}} 
\citep{LSST_SciBook,K10_zphot},
\DESSN\ \citep{DESSN2012}, 
   and
WFIRST\footnote{Wide Field Infrared Space Telescope: \\ {\tt https://wfirst.gsfc.nasa.gov} } 
\citep{Hounsell2018}.

While our main focus is to describe the \DESLOWZ\ simulation of SNe~Ia,
and how a large (${\sim}10^6$ events) simulated bias-correction sample is
used to model biases in the measured distance modulus, 
it is worth noting other applications from the flexibility in \SNANA.
First, these simulations are used to generate 100  data-sized \DESLOWZ\ validation samples
that are processed with the same bias corrections and cosmology analysis used on the data.
This validation test is used to accurately check for $w$-biases at the $\sim0.01$ level,
and to compare the spread in $w$ values with the fitted \unc\  \citep{Brout2018_ANA}.
The validation and bias-correction samples are generated with the same code and options,
but are used for different tasks.
Other applications  include CC simulations for a classification challenge \citep{K10_SNPCC},
CC simulations for a PS1 cosmology analysis using 
photometrically identified SNe~Ia \citep{Jones2017,Jones2018},
simulating the KN search efficiency \citep{Santos2016,Doctor2017},
and making KN discovery predictions for 
11 past, current, and future surveys \citep{ScolnicKN17}.

In this work we describe the simulation from a scientific perspective without
instructions on implementation. For implementation, we refer to the
manual available from the \SNANA\ homepage,
and recommend contacting community members familiar 
with the software. This simulation is possible because of extensive publicly
available resources. When using this simulation in a publication, we 
recommend the added effort of referencing the relevant underlying 
contributions, such as the source of models or data samples used to make templates.

The organization of this paper is as follows.
The \DESLOWZ\ sample is described in \S\ref{sec:data}.
An overview of the simulation method is in \S\ref{sec:overview},
and fake SN light curves overlaid on images is described in \S\ref{sec:fakes}.
Modeling is described in 
\S\ref{sec:model} for  SNe~Ia light curve magnitudes,
\S\ref{sec:noise} for fluxes and \uncs, and 
\S\ref{sec:trigger} for the trigger.
The quality of the simulation is illustrated with data/simulation comparisons
in \S\ref{sec:ovdatasim}, and redshift-dependent $\mu$-biases 
are described in \S\ref{sec:bias}.
We conclude in \S\ref{sec:fin}, and present
additional simulation features in the Appendix.

 \section{Data Samples}
 \label{sec:data}
  
Here we describe the data samples that are simulated for the 
cosmology analysis in \citet{KEY2018} and \citet{Brout2018_ANA}.
After selection,
this sample includes \NDES\ \specy\ confirmed  SNe~Ia from the first three seasons 
(2013 August  through 2016 February) of \DESSN\  \citep{DESY1toY3},
and $\NLOWZ$ \lowz\ ($z<0.1$) SNe~Ia from 
CFA3 \citep{CFA3_Hicken2009},
CFA4 \citep{CFA4_Hicken2012},
and 
CSP \citep{CSP_Contreras2010,CSP_Fol2010}.
This combined sample of  \NTOT\ SNe~Ia is called  ``{\DESLOWZ}.''

The \DESSN\ sample was acquired in rolling search mode using the 570 Megapixel Dark Energy Camera 
(DECam; \citet{DECAM2015})  mounted on the 4-m Blanco telescope at the 
Cerro Tololo Inter-American Observatory (CTIO).
Ten 2.7~deg$^2$ fields were observed in $g,r,i,z$ broadband filters, with a cadence of roughly 1 week in each band.
Defining single-visit depth as the magnitude where the detection efficiency is 50\%,
eight of these fields have an average single-visit depth  of $\sim 23.5$~mag
(hereafter called `shallow' fields), and the remaining two fields have a depth of $\sim 24.5$~mag
(hereafter called  `deep' fields).

SNe~Ia are detected by a difference-imaging pipeline ({\Diff})  described in \citet{Diffimg}, 
and the \spec\ selection is described in \citet{DAndrea2018}. 
The instrumental photometric precision from {\Diff}  is limited at the 2\% level, and therefore
a separate and more accurate ``Scene Model Photometry (\SMP)" pipeline \citep{Brout2018_SMP}
is used to measure the light curve fluxes and \uncs\ for the cosmology analysis.
For each event, {\SMP}  simultaneously fits a $30\times 30$ pixel-grid flux model to each \obs, 
where the model includes a time-independent galaxy flux and a time-dependent
source flux, each convolved with the PSF.

In addition to SN~Ia light curves, the \DESSN\  data  include other `meta-data'  for 
monitoring, calibration (e.g., telescope transmissions) and analysis.
An important meta-data product for simulations is from the 
fluxes of ${\sim}\NFAKE$ `fake'  SN light curves overlaid on the images during the survey 
(\S\ref{sec:fakes}), 
and processed  in real time along with the data to find SN candidates with \Diff.

The \lowz\ sample includes redshifts, light-curve fluxes, flux \uncs,
and filter transmission functions.
The photometry, however, is not from a rolling search but is from follow-up programs
that target SNe~Ia discovered from other search programs such as LOSS \citep{LOSS2013}.
Since the \obs\ information from the search programs is not available,
the resulting \obs\ library is an approximation based on 
several assumptions (\S\ref{sss:cadence_lowz}).

\section{ Overview of Bias Corrections \& Simulation }
\label{sec:overview}

The primary goal of our simulation is to provide inputs to the
`BEAMS with Bias Correction' (BBC) method \citep{BBC}, which is the stage
in our cosmology analysis that produces a bias-corrected SN~Ia Hubble diagram
(\S3.8.1 of \citealt{Brout2018_ANA}). A large simulated bias-correction sample is fit with the
\SALTII\ light curve model, in the same way as for the data, 
to produce three parameters for each event: 
amplitude ($x_0$), stretch ($x_1$), and color ($c$). 
A statistical comparison of the fitted and true parameters is used to
determine a bias correction for each parameter on a 5-dimensional (5D) grid of
$\{z,x_1,c,\alpha,\beta\}$,
where $z$ is the redshift, $x_1$ and $c$ are {\SALTII}-fitted parameters,
and $\alpha$ and $\beta$ are \SALTII\ standardization parameters
(\S\ref{subsec:SALT2par}). 
BBC uses the 5D grid to bias-correct each set of \SALTII\ parameters from the data,
and these corrected parameters are used to determine a bias-corrected distance modulus.

A  schematic illustration of the \SNANA\ simulation is shown in Fig.~\ref{fig:flowChart}.
The left column illustrates the generation of the source 
spectral energy distribution (SED), and astrophysical effects.
These effects include host galaxy extinction, redshifting, cosmological dimming, 
lensing magnification, peculiar velocity, and Milky Way extinction.  
The output of this column is a true magnitude at the top of the atmosphere.
 
The middle column of Fig.~\ref{fig:flowChart} illustrates the instrumental simulation, 
where the true magnitude is converted into an observed 
number of CCD counts, hereafter denoted `flux,'   and the \unc\ on the flux.
The \obs\ information (PSF, sky noise, zero point) and host galaxy
profile are used to compute the Poisson noise. 

The right column in Fig.~\ref{fig:flowChart} illustrates the simulation of the trigger that selects 
events for analysis. Epochs that result in a detection,
which is roughly a $5\sigma$ excess on the subtracted image (\S\ref{subsec:trigger_DESSN}), 
are processed with additional logic  to identify and store `candidates' for analysis.
The candidate logic  specifies  how many detections,  from which band(s) the detection must occur, 
and the minimum time separation between detections.
Finally, the trigger includes a selection function
for the subset of candidates that were {\specy} confirmed.

The noise and trigger models in Fig.~\ref{fig:flowChart} each have inputs
based on analyzing artificial light curves overlaid on CCD images. 
These fakes are described next in \S\ref{sec:fakes}

\section{ Fake SN~Ia Light Curves Overlaid on Images}
\label{sec:fakes}

\begin{figure*}  
\begin{center}
        \includegraphics[angle=-90,scale=0.6]{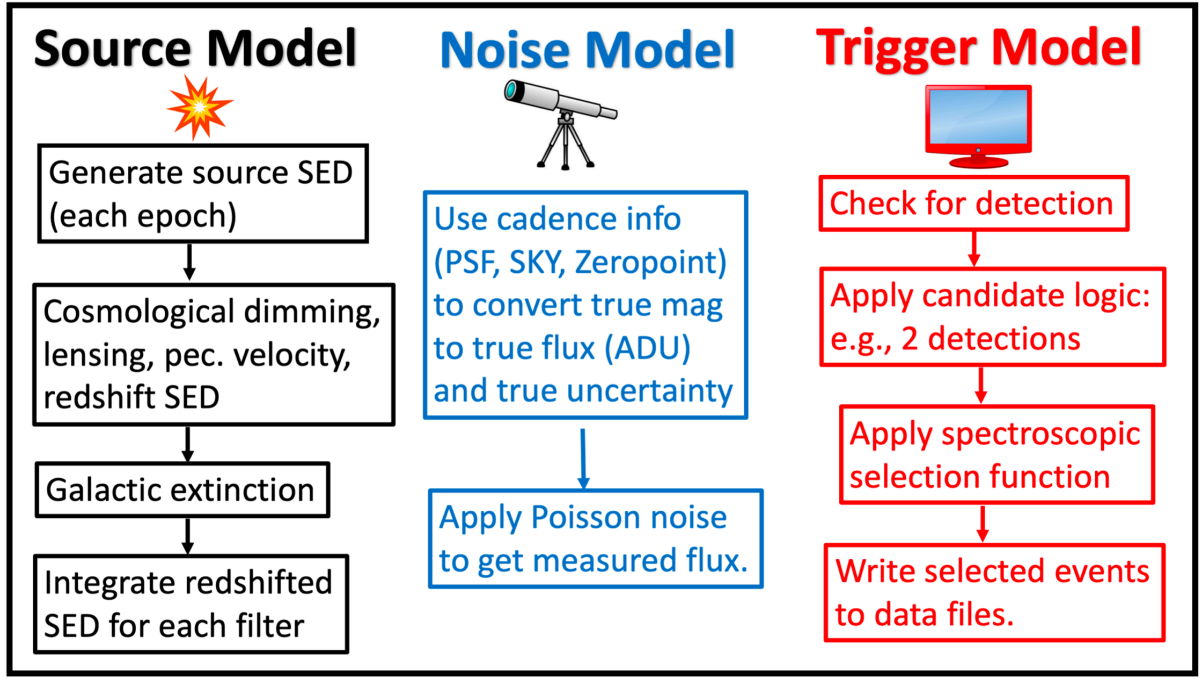}
         \vspace{-0.8in}
   \end{center}
  \caption{ Flow chart of \SNANA\ simulation.}
\label{fig:flowChart}  \end{figure*}

Ideally,  simulated bias corrections would be based on SN~Ia light curve fluxes
overlaid onto CCD images and processed with exactly the same software as the data. 
The CPU resources for so many image-based simulations, however, would be 
enormous. \citet{Diffimg} estimates that \SNANA\ simulations are ${\times}10^5$ 
faster than image-based simulations, while still producing realistic light curve fluxes and \uncs. 
Although we do not perform image-based simulations for bias corrections,
we use image simulations to inform the \SNANA\ simulation. 
Specifically,  \NFAKE\  fake SN light curves were overlaid on \DESSN\ images and 
processed through the same pipelines as the data, including difference-imaging 
(\Diff; \citealt{Diffimg})
and photometry \citep{Brout2018_SMP}  pipelines.
For \Diff, these fakes are used to measure the detection \eff\ versus signal-to-noise ratio (SNR),
which is needed for the trigger model in Fig.~\ref{fig:flowChart}. 
For the photometry pipeline, fakes are used to measure the rms scatter 
between measured and true fluxes, 
and the rms is used to determine scale factors for the SN flux \uncs\
(noise model in  Fig.~\ref{fig:flowChart}).

Prior to the start of DES operations,
the fake light curve fluxes were computed from the \SNANA\ simulation using
the population of stretch and color (\S\ref{subsec:model_SNIa})  from \citet{K13}, 
and  intrinsic scatter (\S\ref{subsec:model_int}) was ignored to simplify the analyses with fakes.
The redshift distribution ($0.1< z < 1.4$) is described by a polynomial
function of redshift, and was tuned to acquire good statistics over the  full redshift range 
and thus span the full range of SN magnitudes.
Each fake location is selected on top of a random galaxy as described in \S\ref{subsec:host}.
The SN model flux is distributed among pixels using the position-dependent PSF,
and the flux in each pixel  includes Poisson fluctuations from the sky background and the source.

The  galaxy occupation fraction was  limited to $\sim 1$\% in each 0.05-wide redshift bin
because the DES-SN search pipeline processed only one set of images, which included fakes, 
and a fake event overlaid on a galaxy  prevents a real transient detection on that galaxy in
the same season, but allows real events in future seasons where the fake is not overlaid.
Once a real transient event is associated with a galaxy,  fakes will not be overlaid on that galaxy.
Since the SMP pipeline performs a global fit to all images, accurate astrometry is needed to 
overlay the fake light curve flux at the same sky location for each exposure.
As described in \citet{Brout2018_SMP},  our astrometric precision results in  
${\sim}0.001$~mag uncertainties for real sources, and thus the
astrometric precision is adequate for the fakes.

Since the goal with fakes is to characterize single-epoch 
features of the CPU intensive image-processing pipelines, 
and to input these features into the much faster \SNANA\ simulation,
the choice of SN light curve model does not matter as long as the 
fake model magnitudes span the same range as the data. 
The resulting \SNANA\ simulation can be used to simulate arbitrary 
light curve models and redshift dependence.
For example, to evaluate systematic \uncs\ in this analysis, 
we simulate SNe~Ia with different models of intrinsic scatter and 
with different populations of stretch and color. 

While this seemingly large sample of SN fakes is used to characterize
image-processing features, these fakes cannot be used  to compute
\mubias\ corrections for two reasons. 
First, the  light curve model used to generate fakes  is deliberately different from 
reality for practical reasons explained above.
Second,  \NFAKE\ fakes is more than an order of magnitude smaller
than what is needed for the bias-correction sample used in the 
BBC method \citep{BBC}. 
In addition,  even if an accurate SN~Ia model were used to generate
fakes, the resulting efficiency and bias corrections would be valid
only for  that particular SN~Ia model, and not applicable to 
other SN~Ia models, nor to  transient models such as  CC SNe or  KNe.

\section{Source Model}
\label{sec:model}

Here we describe the simulation components under ``Source Model" in Fig.~\ref{fig:flowChart}.
This includes the generation of the SN~Ia SED as a function of time,
how the SEDs  are altered as the light travels from the source to Earth,
and how each SED is transformed into a model magnitude above
the atmosphere.

\subsection{ SN~Ia Light Curve Model }
\label{subsec:model_SNIa}

To simulate SNe~Ia, we use the \SALTII\  SED model described in \citet{Guy2010},
and the trained model from the  Joint Lightcurve Analysis \citep{JLA}. 
The underlying model is a
rest-frame SED with wavelengths spanning 2000~\AA\ to 9200~\AA, 
and rest-frame epochs spanning $-20 < \Trest < +50$ days with respect to the 
epoch of peak brightness. 
For each event there are four SN-dependent parameters generated by the simulation:
\begin{enumerate}
    \item time of peak brightness, $t_0$, randomly selected between 2 months before the 
            survey begins and one month after the survey ends.
    \item \SALTII\  color parameter, $c$. 
    \item \SALTII\  stretch parameter, $x_1$. 
   \item CMB frame redshift, $\zcmb$, selected from the rate model in 
           \S\ref{subsec:rate_model}~.
\end{enumerate}            
The rest-frame SED depends on the color and stretch parameters.
For each epoch, the SED 
undergoes cosmological dimming (\S\ref{subsec:model_DL}),
is redshifted (\S\ref{subsec:model_vpec})
to the observer frame,  and finally
multiplied by the filter transmission function to produce generated fluxes and magnitudes.
Wavelength-dependent Milky Way extinction (\S\ref{subsec:model_MWEBV})
is included in the flux-integrals, and thus the generated magnitudes are
top-of-the-atmosphere.
For epochs past 50 days, magnitudes are 
linearly extrapolated  as a function of $\Trest$.
For light curve fitting we use epochs satisfying $-15 < \Trest < 45$~days,
but we simulate epochs outside this $\Trest$ range 
to account for uncertainty in the fitted $t_0$, which increases the true $\Trest$ range.

The \SALTII\ amplitude parameter, $x_0$, is computed 
using the  estimator in \citet{Tripp1998},
\begin{equation}
    \log_{10}(x_0) = -0.4(\mumodel + \muLens - \alpha x_1 + \beta c -\M )~,
    \label{eq:x0}
\end{equation}
where $\mumodel$ is the distance modulus (\S\ref{subsec:model_DL})
which depends on cosmology parameters,
$\muLens$ is due to  lensing magnification (\S\ref{subsec:model_DL}),
$\alpha$ and $\beta$ are \SALTII\ standardization parameters (\S\ref{subsec:SALT2par}),
and $\M = \Mval$ is a reference magnitude.
 It is well known that $\M$ is degenerate with the Hubble constant ($H_0$),
and that  their values have no impact in the SN~Ia analysis of cosmological parameters.
However,  the quality of the simulation depends on predicting accurate observer-frame
magnitudes, and therefore $\M + 5\log_{10}(H_0)$ must be well determined.
The \SALTII\ SED is redshifted using the heliocentric redshift ($\zhel$),
which is transformed from $\zcmb$ using the sky coordinates from the 
\obs\ library. $\zhel$ also includes a random host-galaxy peculiar velocity described in
\S\ref{subsec:model_vpec}.

\subsection{ Intrinsic Scatter }
\label{subsec:model_int}

Before redshifting the \SALTII\ SED, intrinsic scatter is applied as spectral variations to the SED.
To evaluate systematic \uncs\ in the bias corrections, two different models are used that approximately 
span the range of possibilities  in current data samples.
First is the `G10'  model \citep{Guy2010} from the \SALTII\ training process. 
Roughly 75\% of the scatter is coherent among all wavelengths and epochs, 
while the remaining 25\% of the scatter results from color variations
that are not correlated with luminosity.
The second  model, `C11,'  is based on broadband ($UBVRI$) covariances found in \citet{C11}.
Only 25\% of the scatter is coherent, while the remaining scatter results  from color variations.
Details of these models are given in \citet{K13},
and both models result in $0.13$~mag intrinsic scatter on the Hubble 
diagram.\footnote{This 0.13~mag scatter is larger than typical fitted 
   $\sigma_{\rm int}$ values of 0.1~mag because of \SALTII\ model {\uncs}; 
    see \S~7.1 of \citet{BBC} for explanation.}

\subsection{ Global \SALTII\ Model Parameters }
\label{subsec:SALT2par}

While the \SALTII\  SED and color law model parameters from the training process
are fixed for each SN, there are a few parameters that are determined outside the training process.
To simulate validation data samples, the standardization parameters are:
$\alpha=0.15$,
$\beta_{\rm G10} = 3.1$, and 
$\beta_{\rm C11} = 3.8$, 
where G10 and C11 refer to the intrinsic scatter model (\S\ref{subsec:model_int}).
The standardization parameters for the \mubias\ simulations are
defined on a $2\times 2$ grid to enable BBC interpolation of the \mubias\
as a function of $\alpha$ and $\beta$; this is described in \S\ref{sec:bias}.

For the population of stretch ($x_1$) and color ($c$),  we use the asymmetric Gaussian 
parametrization from \citet[hereafter SK16]{SK2016}.  
For \DESSN\ stretch \& color, we use the high-$z$ G10 and C11 rows from  Table~1 of SK16. 
For the \lowz\ sample we use the color population parameters from the
low-$z$ row of Table~1 of SK16. The stretch population is double-peaked,
and we use the parametrization from Appendix~C of \citet{Pantheon}.
While we account for the color and stretch population differences between \lowz\ and \DESSN, 
the redshift dependence of the population has not been quantified
and therefore is not included in our simulations.

\subsection{ Luminosity Distance  and Lensing Magnification }
\label{subsec:model_DL}

The luminosity distance ($D_L$)  for a flat universe 
is computed as
\begin{eqnarray}
   D_L     & =  & (1+\zhel) \frac{c}{H_0} \int_0^{\zcmb} dz/E(z)  \\
   E(z)   &  =  &  \left[ \OL(1+z)^{3(1+w) } + \OM(1+z)^3  \right]^{1/2}~,  \nonumber
\end{eqnarray}
where $\OM$ is today's matter density, $\OL$ is today's dark energy density,
and $w$ is the dark energy equation of state parameter.
Note that both the CMB and heliocentric redshifts are used to compute $D_L$,
but the $1+\zhel$ pre-factor is an approximation:
the exact pre-factor is $1+\zobs$, where $\zobs$ is the measured redshift  \citep{Davis2011}.
However, we do not simulate Earth's motion around our Sun, nor the local
SN motion within its host galaxy, and therefore $\zobs=\zhel$. 
The error on $D_L$ from ignoring local motions is less than $10^{-3}$.
To compute $E(z)$ in our simulations we set 
$H_0=70~{\rm km/s}$,\footnote{This $H_0$ value was used in the
   SN~Ia model training, which determines the absolute brightness $\M$, and 
   therefore the simulated $H_0$ should not be updated with more recent measurements.}
$\OL=0.7$, $\OM=0.3$ and $w=-1$.
The distance modulus ($\mumodel$ in Eq.~\ref{eq:x0})
is defined as  $\mumodel = 5\log_{10}(D_L/10~{\rm pc})$.

Weak lensing effects are described by the $\muLens$ term in Eq.~\ref{eq:x0}, and
modeled as follows:
\begin{itemize}
   \item  $0.4 < z< 1.4$ : 
         A convergence ($\kappa$) distribution is determined
         from a  900~deg$^2$ patch of the 
         MICECAT $N$-body simulation \citep{MICE2015}.\footnote{https://cosmohub.pic.es}
         Galaxies are from a halo occupation distribution and a 
         sub-halo abundance matching technique \citep{MOCKGAL}.
	The lensing distribution is determined from $\muLens = 5\log_{10}(1-\kappa)$ 
	(shear contribution is negligible and ignored).
   \item $z < 0.4$ : 
      	 Using the MICECAT approach above to determine  $\muLens$     	
         at $z=0.4$ ($\muLensA$),  the lensing at lower redshifts is computed as
          \begin{equation} \muLensz = \muLensA \times z/0.4~.  \end{equation}
        As a crosscheck, this $z$-scale approximation works well within the 
        MICECAT redshift range: e.g., $\muLensA \simeq \muLensB \times (0.4/0.6) $. 
\end{itemize}
The root mean square (rms) scatter in the model $\muLens$ distribution 
is approximately $0.05\times z$. For systematic studies, the simulation includes 
an option to scale the width of the distribution to increase or decrease the scatter.

To properly select from the asymmetric $\muLens$ distribution,
instead of a Gaussian approximation,
the lensing magnification probability is defined as a 2-dimensional
function of  redshift and $\muLens$.
For each simulated redshift, a random $\muLens$ is selected
from the $\muLens$ probability distribution. 
While our lensing model accounts for large scale structure on average,
it does not account for correlations between events with small angular separations.

\subsection{ Peculiar Velocity and Observed Redshift }
\label{subsec:model_vpec}

The generated CMB-frame redshift, $\zcmb$, is transformed to the heliocentric frame, 
$\zhel$, using the sky coordinates from the \obs\ library.
The redshift observed in the heliocentric frame is 
\begin{eqnarray}
   \zhelObs & = & (1+\zhel)(1 + \vpec/c - \vcor/c) -1 \\
                  &   & + \dzNoise  \nonumber \\
                   & = & (1+\zhel)(1 - \verr/c) -1 \label{eq:vpec} \\
                    & & + \dzNoise~,   \nonumber
\end{eqnarray}
where $\vpec$ is a peculiar velocity randomly chosen from a Gaussian profile with 
$\sigmavpec = \SIGMAVPEC$~km/sec, and $\dzNoise$ is a measurement error.
For \DESSN\ and \lowz, $\dzNoise$ is drawn from a Gaussian with   $\sigz=\sigzOZDES$.

While the peculiar velocity model is the same for \lowz\ and \DESSN,
corrections are modeled only for the \lowz\ sample.
The simulated \lowz\ correction simply reduces the $\vpec$ scatter
without  applying real corrections.
Following the Pantheon analysis \citep{Pantheon},
$\vcor = \vpec + \verr$ where $\verr$ is a randomly selected error
from a Gaussian profile with a $\SIGMAVCOR$~km/sec sigma.
Finally, $\zhelObs$ is transformed back to the CMB frame redshift, $\zcmbObs$.
Peculiar velocity corrections for \DESSN\ can be computed in principle,
but such corrections on a high-redshift sample are negligible and were thus ignored. 

\subsection{Galactic Extinction}
\label{subsec:model_MWEBV}

For each simulated event the Galactic extinction parameter, $E(B-V)$, 
is computed from the maps in \citet{SFD98}.
Following a stellar analysis from SDSS \citep{SF2011},
we scale the $E(B-V)$ values by 0.86.
We assume the reddening law derived in \citet{Fitz1999}, 
and with $A_V$ defined as the extinction at 5500~\AA,
$R_V\equiv A_V/E(B-V) =3.1$.

\subsection{Volumetric Rate Model}
\label{subsec:rate_model}

The redshift distribution of SNe~Ia is generated from a co-moving
volumetric rate, $R(z)$,  
measured by SNLS \citep{Perrett2012}:
\begin{equation}
    R(z)   =  1.75\times 10^{-5} (1+z)^{2.11} ~{\rm yr}^{-1}{\rm Mpc}^{-3} ~, \\
\end{equation}
and is valid up to redshift $z < 1$.

\section{Modeling Flux and Noise} 
\label{sec:noise}

Here we describe the simulation components under `Noise Model'  in Fig.~\ref{fig:flowChart}.
There are two steps needed to  simulate flux and noise.
First, an \obs\ library is needed to characterize observing conditions
(\S\ref{subsec:cadence}).
Next, each true model magnitude is converted into a measured
flux (CCD counts) and \unc\ (\S\ref{subsec:model_fluxErr}).

\subsection{Observation Library}
\label{subsec:cadence}


An \obs\ library is a collection of sky locations, 
each specified by right ascension (RA) and declination (Dec),
along with a list of \obss\ for each location.
For a small survey area, a single sky location may be adequate, 
particularly for making forecasts. For a proper simulation, however, many 
sky locations should be used with either random sampling or a grid.
For \DESSN\ we use $\sim 10^4$ random sky locations covering 27~deg$^2$,
which averages over density fluctuations to achieve a representative sample
for a homogeneous universe.
For simulations with more than $10^4$ generated events, the library 
sky locations and \obss\ are re-used with SNe that have a different set of 
randomly chosen properties.
Since $\sim 1$\% of the \DESSN\ events occur in the overlap between two adjacent fields,
and thus have double the number of \obss, 
the simulation includes a mechanism to handle overlapping fields.

The exposure information for each sky location is defined as follows:
\begin{itemize}
  \item \MJD\ is the modified Julian date.
  \item {\tt FILTER} is the filter passband.
  \item \GAIN\ is the number of photoelectrons per 
            ADU\footnote{ADU: Analog to Digital Unit.}.
  \item \SKYSIG\ is the sky noise, including read noise.
  \item ${\PSFSIG}  =  \sqrt{ \NEA\ /4\pi  }$ is an effective Gaussian $\sigma$ for the PSF, 
  	and \NEA\ is the noise-equivalent area defined as         
    \begin{equation}
      \NEA\  =   \left[ 2\pi\int [{\rm PSF}(r)]^2 r dr  \right]^{-1} ~. 
        \label{eq:NEA}
   \end{equation}
   For a PSF-fitted flux, the fitted flux variance is the sky noise per pixel
   multiplied by \NEA.
  \item \ZPTADU\ is the zero point (ADU), and includes 
              telescope and atmospheric transmission.
\end{itemize}

Many of the \DESSN\ visits include multiple exposures:
two $z$-band exposures in each of the 8 shallow 
fields,\footnote{shallow $g,r,i$ include one exposure per visit.}
and the 2 deep fields include 3, 3, 5, 11 exposures for $g,r,i,z$-bands, respectively.
During the survey, \Diff\ performs the search on co-added exposures.
In the analysis, \SMP\ determines the flux for each individual exposure,
and the fluxes are co-added at the catalog level.
The co-adding for both \Diff\ and \SMP\ are treated the same in the
simulation by co-adding the \obs\ library information as follows:

\newcommand{\Nexp}{N_{\rm expose}}
\begin{eqnarray}
    \MJD\    & = & \left[ \sum \MJD_i \right] / \Nexp   \nonumber \\
    \SKYSIG\   & = & \sqrt{ \sum \SKYSIG_i^2 } \nonumber \\
    \PSFSIG       & = & \left[ \sum {\PSFSIG}_i \right] / \Nexp \nonumber \\
    \ZPTADU\   & = & 2.5 \times \log_{10} \left[ \sum 10^{(0.4 \cdot \ZPTADU_i)} \right] ~,
\end{eqnarray}
where $\Nexp$ is the number of exposures and each sum includes $i=1, \Nexp$.
\ZPTADU\ is an approximation assuming the same \GAIN\ for each exposure;
the DES \GAIN\ variations are a few percent.

The randomly selected time of peak brightness ($t_0$, \S\ref{subsec:model_SNIa}),
along with the light-curve time window,
determine the {\MJD}-overlap in the \obs\ sequence.

\subsubsection{ Observation Library for Low-$z$ Sample }
\label{sss:cadence_lowz}

The \lowz\ sample does not include the \obs\ properties (PSF, sky noise, zero point)
from their image-processing pipelines.
Therefore we construct an approximate library from the \lowz\  light curves,
using their sky locations, \obs\ dates, and SNR.
There is not enough light curve information to uniquely
determine the \obs\ properties, and therefore we use three assumptions: 
(1) fix each \GAIN\ to unity,
(2) fix each  PSF to $1^{\prime\prime}$ (FWHM), and
(3) use a previously determined set of broadband sky magnitudes,
and interpolate the sky magnitude to the central wavelength of each 
simulated filter. For ground-based surveys we use the average sky mags in 
$ugrizY$ passbands from a simulation of LSST \citep{LSST_OPSIM}.
The \ZPTADU\ parameter is adjusted numerically so that the
calculated SNR matches the observed SNR.

Another subtlety is that the \lowz\ sample was collected over decades,
and thus for a randomly selected explosion time there is little chance 
that the simulated light curve would overlap the \obs\ dates.
To generate \lowz\ light curves more efficiently, the measured time
of peak brightness ($t_0$)  for each light curve is used for the
corresponding sky location, thus ensuring a light curve will be generated.
Other SN properties (redshift, color, stretch, intrinsic scatter) 
are randomly selected in the same way as for the \DESSN\ simulation.

\subsection{Host Galaxy Model}
\label{subsec:host}

Host galaxies are used for two purposes in the 
\SNANA\ simulations of  \DESSN. 
First, fakes are generated to be overlaid on  top of galaxies in real CCD images.
Second, to simulate bias corrections and validation samples in the analysis,
the local surface brightness from the host is used to 
add Poisson noise and anomalous scatter  (Fig.~\ref{fig:fluxerrscale})
in the light curve fluxes.
We do not simulate \lowz\ host galaxies because the cadence library
is constructed from observed SNR that should already include Poisson noise from
the host. While anomalous scatter in the \lowz\ sample may be present,
the local surface brightness information is not available to study 
this effect.\footnote{It would be a valuable  community contribution to
    use public survey data  (e.g., PS1, SDSS, DES) and determine the local surface brightness 
    for  each \lowz\ event.}

For \DESSN\ the {\SNANA} simulation uses a host galaxy library (\HOSTLIB), 
where each galaxy is described by 
(1) heliocentric redshift, $\zhelHOST$,
(2) coordinates of the galaxy center,
(3) observer-frame magnitudes in the survey bandpasses, and
(4) \Sersic\ profile with index $n=0.5$ (Gaussian), 
 and half-light radii along the major and semi-major axes.
The \HOSTLIB\ can be created from data or from an astrophysical simulation.
Our \DESSN\ simulation uses a galaxy catalog derived from the 
DES science verification (SV) data, as described in \citet{Gupta2016}.
Eventually this galaxy catalog will be updated
using a much deeper co-add from the full DES data set.

\newcommand{\muphot}{\mu_{\rm phot}}

There are two caveats about this \HOSTLIB.
First, $\zhelHOST$ are photometric redshifts (photo-$z$). 
Extreme photo-$z$ outliers are rejected by requiring the absolute $r$ and $i$ band magnitudes ($M_{r,i}$)
to satisfy $-23 < M_{r,i} < -16$, where $M_{r,i} = m_{r,i} - \muphot$, $m_{r,i}$ are the observed magnitues,
and $\muphot$ is the distance modulus computed from the photo-$z$.
The second caveat is that the measured half-light radii were scaled by a factor of 0.8 to obtain better 
data-simulation agreement in the surface brightness distribution (\S\ref{sec:ovdatasim}).

To generate fakes to overlay on images, each  SN was associated with a random 
\HOSTLIB\ event satisfying 
\begin{equation}
   \vert \zhelSN - \zhelHOST \vert < 0.01 + 0.05\zSN~,
     \label{eq:zhost_match}
\end{equation}
where $\zhelSN$ and $\zhelHOST$ are the heliocentric redshifts for the
SN and host galaxy, respectively. 
The SN redshift is updated to $\zhelHOST$,
the CMB-frame redshift ($\zcmb$) is computed from $\zhelHOST$, and the resulting $\zcmb$
is used to update the distance modulus and light curve magnitudes.
To avoid multiple fakes around a single galaxy, each \HOSTLIB\ event can be 
used only once.
The SN coordinates are chosen near the host, weighted by the \Sersic\ profile.

To simulate samples for the analysis, the redshift matching between the SN and the host 
is the same as for fakes (Eq.~\ref{eq:zhost_match}). However, the generated SN 
redshift (from rate model) and its coordinates (from cadence library) are preserved.
A random location near the host is selected from the \Sersic\ profile,
and is used to determine the local surface brightness and to add Poisson 
noise to the light curves. 
The Poisson noise variance is computed by integrating the
host-galaxy flux over the noise equivalent area (Eq.~\ref{eq:NEA}).
In this implementation of the \DESSN\ simulation, the host-galaxy spatial distribution
is homogeneous on all scales. Large-scale structure can be incorporated
as explained in the Appendix.

\subsection{Converting True Magnitudes into  Measured  \\ Fluxes \& Uncertainties }
\label{subsec:model_fluxErr}

Here we describe how a true source magnitude at the top of the atmosphere, 
$\mtrue$, is used to determine the instrumental flux and its uncertainty. 
The flux unit for this discussion is photoelectrons,
but the simulation uses the \GAIN\ to properly digitize the signals in ADU.

The true flux is given by 
\begin{equation}
   \Ftrue = 10^{ 0.4(\mtrue-\ZPTpe) }~,
   \label{eq:Ftrue}
\end{equation}
where $\ZPTpe = \ZPTADU + 2.5\log_{10}(\GAIN)$ is the zero point in units
of photoelectrons.

The true Poisson noise for the measured flux is given by
\begin{equation}
  \sigFtrue^2  =   [  \Ftrue + (\NEA\cdot b)  +  \sigHOST^2  ]  \ERRSCALESIM^2  ~,
             \label{eq:sigF} 
\end{equation}
where
\begin{itemize}
  \item $\Ftrue$ is the true flux;
  \item \NEA\ is the noise equivalent area (Eq.~\ref{eq:NEA});
  \item $b$ is the background per unit area 
      (includes sky and CCD read noise);
  \item $\sigHOST$ is Poisson noise from the underlying host galaxy (\S\ref{subsec:host});
  \item $\ERRSCALESIM$ is an empirically determined scale (\S\ref{subsec:ERRSCALE})
         that increases both the flux scatter and measured \unc.
\end{itemize}
\NEA, $b$, and $\PSFSIG$  are obtained from the \obs\ library (\S\ref{subsec:cadence}).
$\ERRSCALESIM$ is determined from analyzing the fakes (\S\ref{subsec:ERRSCALE}),
and characterizes subtle scene model photometry (\SMP) behavior that
cannot be computed from first principles, mainly the anomalous flux scatter from bright galaxies.
Because of the large number of reference images used in \SMP,
we do not include an explicit template noise term.

For PSF-fitted fluxes, the noise estimate in Eq.~\ref{eq:sigF} is an approximation that 
is more accurate for sky-dominated noise, or as $\Ftrue/(\NEA\cdot b)$ becomes smaller. 
In principle Eq.~\ref{eq:sigF} is also accurate for bright events with high SNR, 
but brighter SNe are associated with brighter galaxies that introduce anomalous 
flux scatter.  
In \S\ref{subsec:ERRSCALE} below, we use fakes to show how the simulated flux \uncs\ are corrected
for anomalous scatter so that the \uncs\ are accurate over the full range of SNR.

The true \unc\ ($\sigFtrue$) is used to select a random fluctuation 
on the true flux ($\Ftrue$), resulting in the observed flux, $F$.
The measured \unc\ for data is not the true \unc, but  rather an approximation based
on the observed flux. In the simulation, the measured \unc, $\sigF$,
is computed from the observed flux 
by substituting $\Ftrue \to F$ in Eq.~\ref{eq:sigF}: 
\begin{eqnarray}
    \sigF & = & \sqrt{\sigFtrue^2 + (F - \Ftrue) }   ~~~~(F>0)~, \\
    \sigF & = & \sqrt{\sigFtrue^2 - \Ftrue }   ~~~~~~~~~~~(F<0) ~.
\end{eqnarray}
In the case where $F<0$ due to a sky noise fluctuation, 
the measured \unc\ is not reduced (relative to $F=0$)
because $\sigF$ is dominated by sky noise which is determined 
from a CCD region well away from the SN.

\subsection{ Determining The Flux-Uncertainty Scale ($\ERRSCALESIM$) }
\label{subsec:ERRSCALE}

An accurate description of the \unc\ is important in order to model
selection cuts on quantities related to SNR and chi-squared from light curve fitting.
With $\ERRSCALESIM=1$, the calculated flux \unc, $\sigFtrue$ 
in Eq.~\ref{eq:sigF}, is an approximation for PSF-fitting, and it 
does not account for all of the details in the \SMP\ pipeline.
We correct the simulated \unc\ to match the observed flux scatter in the fakes,
which we interpret to be the true scatter in the data. 
The \unc\ correction, $\ERRSCALESIM$, is defined as
\begin{equation}
    \ERRSCALESIM(\Ovec) = 
       \frac{ {\rm rms}[ (\Ftrue-\FSMP)/\sigFP ]_{\rm fake} }{ {\rm rms}[ (\Ftrue-\FSIM)/\sigFP ]_{\rm sim} }~,
       \label{eq:errScale_sim}
\end{equation}
where $\Ftrue$ is the true flux, 
$\FSMP$ is the fake flux determined by \SMP, and
$\FSIM = \Ftrue + \GaussSigF$ is the simulated flux with $\ERRSCALESIM=1$.

The $\sigFP$ term in both denominators is a common reference 
so that the $\Delta F/\sigFP$ ratios in Eq.~\ref{eq:errScale_sim} are $\sim$unity, 
which significantly improves the sensitivity in measuring the $\ERRSCALESIM$ map. 
$\sigFP$ is the naively expected {\unc} computed from Eq.~\ref{eq:sigF} with
$\ERRSCALESIM=1$, $\Ftrue \to F$, and $\sigHOST$ computed using the
approximation of a constant local surface brightness magnitude 
over the entire noise-equivalent area. This $\sigHOST$ approximation 
can be used with photometry that does not include a detailed
model of the host galaxy profile,
and simulation tests have shown that this approximation
does not degrade the determination of $\ERRSCALESIM(\Ovec)$.

\begin{figure} 
\centering
\begin{center}
        \includegraphics[scale=0.4]{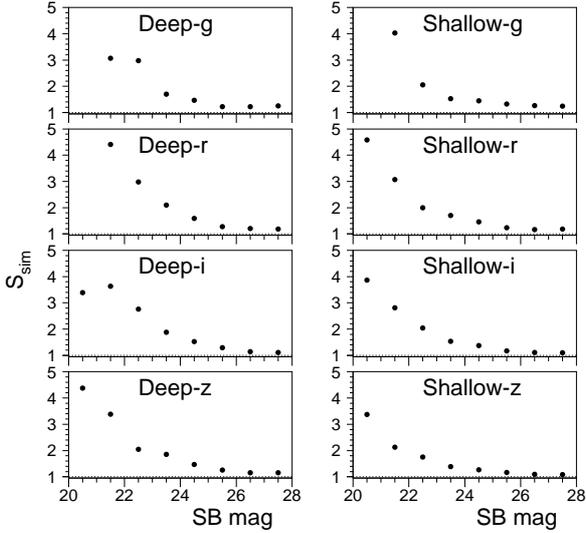}
   \end{center}
  \caption{
    Simulated \unc\ scale, $\ERRSCALESIM$, as a function of
    local surface brightness mag (SB mag). Each panel indicates
    the set of fields and passband. 
    Left panels are for the deep SN fields (depth per visit $\sim 24.5$);
    right panels are for shallow SN fields (depth per visit $\sim 23.5$).
  }
\label{fig:fluxerrscale}  \end{figure}

The numerator includes information from the fakes and \SMP\ pipeline.
The argument $\Ovec$ indicates an arbitrary dependence on  observed image properties.
For the \DESLOWZ\ analysis we use a 1-dimensional map with $\Ovec = \{ \mSB \}$,
where $\mSB$ is the local surface brightness magnitude.
Before determining $\ERRSCALESIM$,  it is important that the simulated distributions in
redshift, color, and stretch  (\S\ref{subsec:model_SNIa})  are tuned to match the distributions for the fakes.
After this tuning, $\ERRSCALESIM$ versus $\mSB$ is shown in Fig.~\ref{fig:fluxerrscale}.
For $\mSB$ values outside the defined range of the map,
$\ERRSCALESIM$ is computed from the closest $\mSB$ value in the map.
This $\mSB$-dependence has been seen previously in the
difference-imaging pipeline \citep{Diffimg,Doctor2017}, 
and it persists in the \SMP\ photometry.
After applying the corrections in Fig.~\ref{fig:fluxerrscale}, the
flux \uncs\ for the fakes and simulations agree to within 5\% over the entire $\mSB$ range. 

\begin{figure}
\centering
\begin{center}
        \includegraphics[scale=0.4]{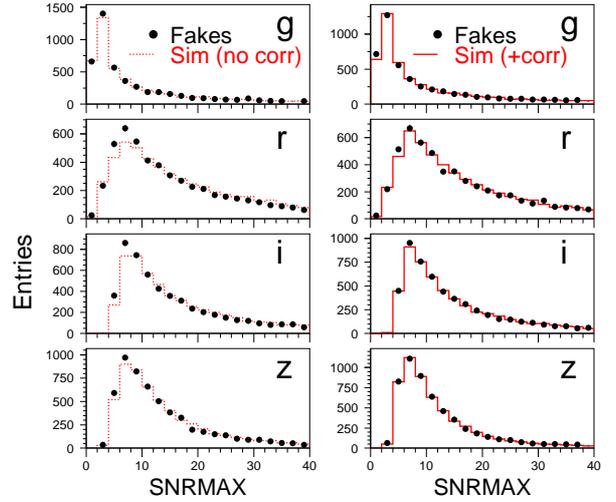}
   \end{center}
  \caption{
    Distribution of maximum SNR in the DES band labeled on each panel.
    Filled circles are for the fakes processed through the \SMP\ pipeline;
    histogram is the simulation.    
    Left panels are before applying the $\ERRSCALESIM$ scale;
    right panels are after applying the  $\ERRSCALESIM$ scale from 
    Fig.~\ref{fig:fluxerrscale}.
  }
\label{fig:ovfakesim_snrmax}  \end{figure}

The impact of the \unc\ corrections is shown in Fig.~\ref{fig:ovfakesim_snrmax},
which compares the maximum SNR distribution in each band 
for fakes and the simulation. Compared to simulations with no correction,
simulations with corrections show much better agreement with the fakes.

While Eq.~\ref{eq:errScale_sim} describes the simulated correction,
there is an analogous correction for the 
data \unc\  produced  by \SMP: $\sigSMP \to \sigSMP \times \ERRSCALESMP$, where
\begin{equation}
    \ERRSCALESMP(\Ovec) = 
       \frac{ {\rm rms}[ (\Ftrue-\FSMP)/\sigFP ]_{\rm fake} }{ {\langle\sigSMP/\sigFP\rangle}_{\rm fake} }~.
       \label{eq:errScale_data}    
\end{equation}
The observed scatter in the fakes is a common reference for both the data and simulations,
and therefore the numerator (Eq.~\ref{eq:errScale_data}) is the same as for the simulated 
correction (Eq.~\ref{eq:errScale_sim}).
The denominator, ${\langle\sigSMP/\sigFP\rangle}_{\rm fake}$, 
specifies an average within each $\Ovec$ bin.
This $\ERRSCALESMP$ correction is applied to the data \uncs, including fakes, 
while $\ERRSCALESIM$ is applied to the simulated noise and \unc. 
More details of $\ERRSCALESMP$ are given in  \citet{Brout2018_SMP}.

\section{ Trigger Model}
\label{sec:trigger}

Here we describe the simulation components under `Trigger Model' in Fig.~\ref{fig:flowChart}. 
Ideally, every DECam pixel would be continuously monitored for transient activity.
However, storing light curves near every pixel is impractical with today's computing,
and therefore a `trigger' is used to select candidates for analysis. Here we describe
the trigger simulation for the \DESSN\ and \lowz\ samples. 
For a general survey, the simulation of the trigger consists of three stages: 
1) detecting PSF-shaped objects above threshold, 
2) matching multiple objects, from different bands and nights, to form candidates, and 
3) selection for \spec\  classification.
All three stages are modeled for \DESSN.
For {\lowz},  however, we do not have information to simulate the first 
two trigger stages and  
therefore all three trigger stages are empirically combined in the third stage.

The total \eff\ ($\EFFTOT$) can be described by
\begin{equation}
   \EFFTOT = \EFFSNR \times \EFFspec~,
\end{equation}   
where $\EFFSNR$ includes the first two trigger stages and depends
on the SNR for each epoch ($\vecSNR$), and $\EFFspec$ describes
the \spec\ selection in the third stage. 
We do not explicitly define $\EFFSNR$, but instead model the
efficiency vs. SNR for each epoch. $\EFFspec$, however, 
is explicitly described by a smooth function of magnitude at peak brightness.
Another subtlety here is that $\EFFSNR$ is valid for arbitrary 
transient source models, while $\EFFspec$ is valid only for SNe~Ia
and only if the first two trigger stages are satisfied.

\subsection{ \DESSN\ Trigger}
\label{subsec:trigger_DESSN}

For the first trigger stage,  fakes are used to characterize the detection \eff\ versus SNR
in each filter, as shown in Fig.~8 of \citet{Diffimg}. 
The \eff\ reaches 50\% around SNR$\sim 5$.
Since these \eff\ curves are intended for simulations, we do not use the measured
SNR, but instead the fake SNR is calculated from the true flux (Eq.~\ref{eq:Ftrue})
and noise (Eq.~\ref{eq:sigF}) with $\ERRSCALESIM=1$.
These \eff\ curves are therefore determined as a function of 
a calculated SNR quantity that is calculated in exactly the same way in the simulation.

In the second trigger stage, a candidate requires two detections on separate nights
within 30 days.
Thus a single-night detection in all four bands ($g,r,i,z$) will not trigger a candidate.
However, a single-band detection on two separate nights will trigger a candidate.

The third trigger stage, \spec\ selection \eff\ ($\EFFspec$), is the most subtle. 
While the selection algorithm was designed to exclude human decisions as much 
as possible \citep{DAndrea2018}, we are not able to simulate the selection algorithm 
because we have 
eight frequently used telescopes, 
inefficiencies due to weather and scheduling, 
spectral classification uncertainty, 
and a small amount of human decision making.

Ideally we would compute $\EFFspec$ as a ratio of \specy\ confirmed
events (numerator)  to photometrically identified events (denominator). 
A data-derived $\EFFspec$ analysis is under development and described in \citet{DAndrea2018}, 
but here we use simulations to predict the denominator. 
A caveat is that a simulation used to determine $\EFFspec$ needs the
population parameters for stretch and color (\S\ref{subsec:SALT2par}), 
which is determined  from simulations that already include $\EFFspec$.
Rather than performing an iterative procedure with \DESSN\ data,
we use the population parameters from external data sets as described in SK16,
who show that varying the external $\EFFspec$ functions has a negligible effect on
the population parameters.

\begin{figure}
\centering
\begin{center}
        \includegraphics[scale=0.35]{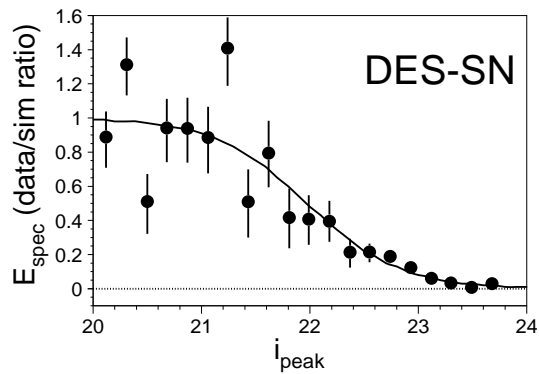}
   \end{center}
  \caption{
    $\EFFspec$ vs. $\ipeak$ for \DESSN\ sample.  
    Filled circles are data/sim ratios, 
    with error bars from Poisson uncertainties on the best-fit model curve.
    The simulation includes the first two trigger stages ($\EFFspec = 1$), 
    uses the G10 scatter model, and is scaled to match the data statistics for the 
    brightest events.
    Smooth solid curve is a fit that defines $\EFFspec$ in the simulation,
    and max $\EFFspec=1$ is an arbitrary normalization.
  }
\label{fig:effspec_DES}  \end{figure}

Without a well-defined algorithm to compute $\EFFspec$,
we use an empirical model where $\EFFspec$ depends on
the $i$-band magnitude at the epoch of peak brightness, $\ipeak$.
The basic idea is to compare the $\ipeak$ distribution between data and a
simulated sample passing  the first two trigger stages (i.e., with $\EFFspec=1$).
We define $\EFFspec(\ipeak)$ to be a smooth curve fit to the
data/sim ratio as a function of $\ipeak$ (solid curve in Fig.~\ref{fig:effspec_DES}),
where  $\ipeak$ is computed from the best-fit \SALTII\ model.
The data/sim ratio is fit to a sigmoid function,
\begin{equation}
   \EFFspec(\ipeak) = s_0 [1 + e^{(s_1\ipeak-s_2)} ]^{-1}~,
\end{equation}
where $s_0,s_1,s_2$ are floated parameters determined with {\tt emcee} \citep{EMCEE2013} 
and the data uncertainties are modeled using a Poisson distribution.
For the cosmology analysis, 
$\EFFspec$ can be arbitrarily scaled (bounded between 0 and 1)
without affecting the \mubias\ determination, and thus to generate
events most efficiently we have scaled $\EFFspec$ to have a maximum  efficiency of 1.

There is a subtle caveat in the \Diff\ trigger modeling related to bright galaxies.
As illustrated in Figure~7 of \citet{Doctor2017},  image-subtraction artifacts result in
an anomalous decrease in detection \eff\ as the local surface brightness increases.
Here the term `anomalous'  indicates an \eff\ loss that is much greater
than expected from the increased Poisson noise from the host galaxy.
While Fig.~\ref{fig:fluxerrscale} shows how the \SNANA\ simulation models
anomalous scatter, the simulation does not model the anomalous
detection inefficiency.
Studies with fakes have shown that this bright-galaxy anomaly 
does not reduce the trigger \eff\ for nearby SNe~Ia on bright
galaxies. The reason is that there are a few dozen opportunities 
to acquire detections, and it is very unlikely to fail the
2-detection trigger requirement.

\subsection{ Low-$z$ Trigger }
\label{subsec:trigger_LOWZ}

As explained in \citet{JLA} and \citet{Scolnic2014cosmo}, there is evidence that the \lowz\ search
is magnitude limited because of the decreasing number of events with redshift,
and because  higher redshift events are bluer. On the other hand, many \lowz\ searches
target a specific list of galaxies, suggesting a volume-limited sample.
We therefore simulate both assumptions for evaluating systematic \uncs.  

For the magnitude-limited assumption, we incorporate all trigger stages into 
a single $\EFFspec$ function of $B$-band magnitude at the time of peak 
brightness ($\Bpeak$).  Following the recipe for the \DESSN\ simulation,
we simulate a \lowz\ sample with $\EFFSNR=1$ and  define $\EFFspec$ to be 
the data/sim ratio vs. $\Bpeak$ (Fig.~\ref{fig:effspec_LOWZ}).
The fitted $\Bpeak$  function is a one-sided Gaussian as described in 
Appendix~C of \cite{Pantheon}.
Describing $\EFFspec$ as a function of $V$ or $R$ band also
works well, so the choice of $B$ band is arbitrary.

For the volume-limited assumption, which is used as a systematic \unc\ in  \citet{Brout2018_ANA},
we set $\EFFTOT=1$ and interpret the redshift evolution of stretch and color  
to be astrophysical effects instead of artifacts from Malmquist bias.
To match the \lowz\ data,  the \lowz\ simulation is tuned using
redshift-dependent stretch and color populations:
$x_1 \to x_1 + 25z$ and  $c \to c - 1z$.
There is no physical motivation for this redshift dependence,
and therefore this is a  conservative assumption for the systematic \unc.

\begin{figure} 
\centering
\begin{center}
        \includegraphics[scale=0.38]{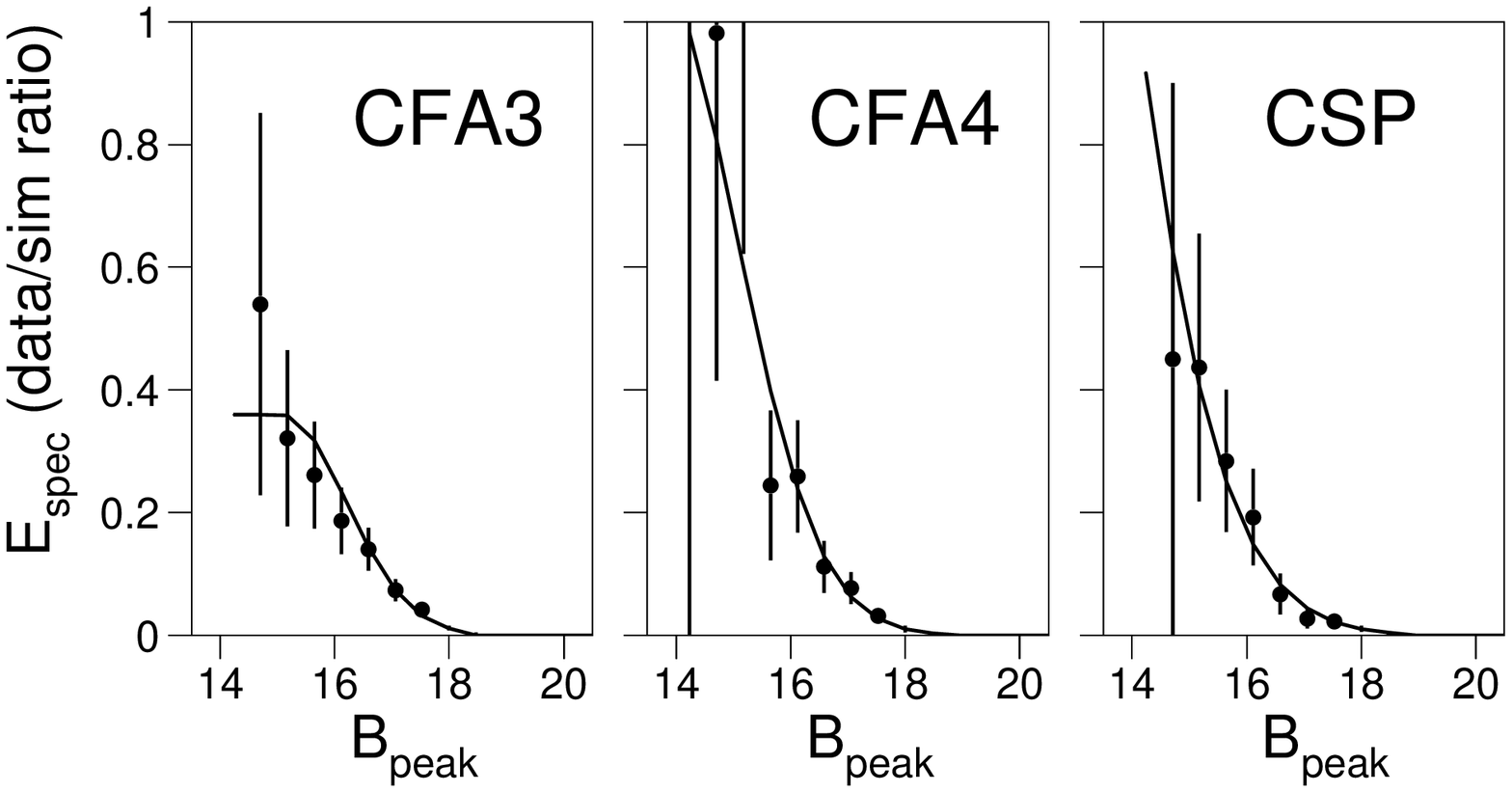}
   \end{center}
  \caption{
    $\EFFspec$ vs. $\Bpeak$ for each \lowz\ sample.
    Filled circles are the data/sim ratio with $\EFFspec = 1$ in the simulation. 
    Smooth curve is a fit that defines $\EFFspec$ in the simulation.
  }
\label{fig:effspec_LOWZ}  \end{figure}

\section{ Comparing Data \& Simulations }
\label{sec:ovdatasim}

Here we qualitatively validate the simulations by comparing simulated distributions with data. 
While we do not quantify the data-simulation agreement here (e.g., via $\chi^2$), 
such quantitative comparisons are used to assess systematic \uncs\ in  \citet{Brout2018_ANA}. 
To limit statistical \uncs\ in these comparisons, very large simulations are generated and the 
distributions are scaled to match the statistics of the data.
Recall that the tuned distributions are $\EFFspec(\ipeak)$ and the populations for stretch and color;
all other inputs to the simulation are from measurements.

\begin{figure*}
\begin{center}
\includegraphics[scale=0.38]{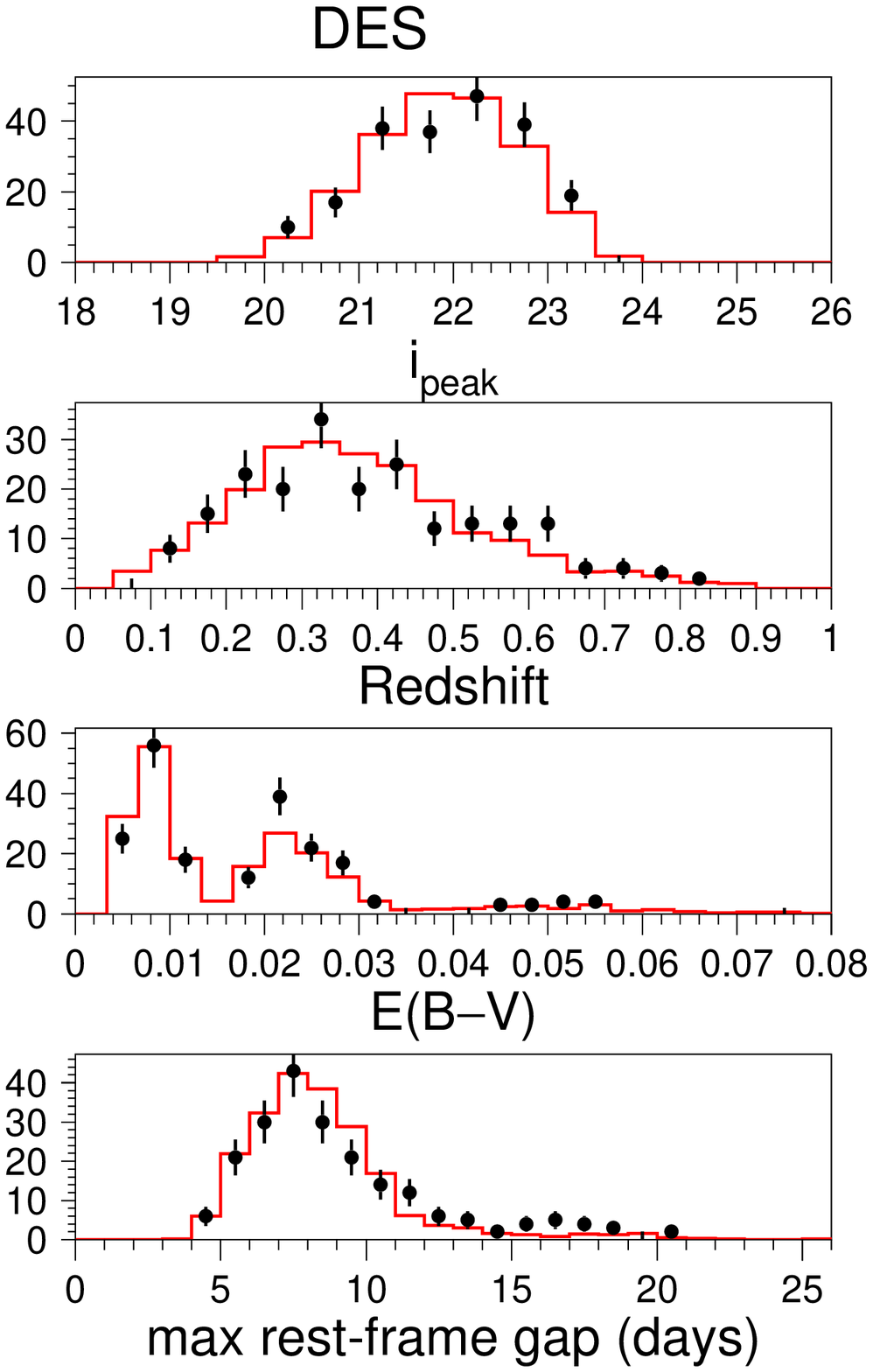}
\includegraphics[scale=0.38]{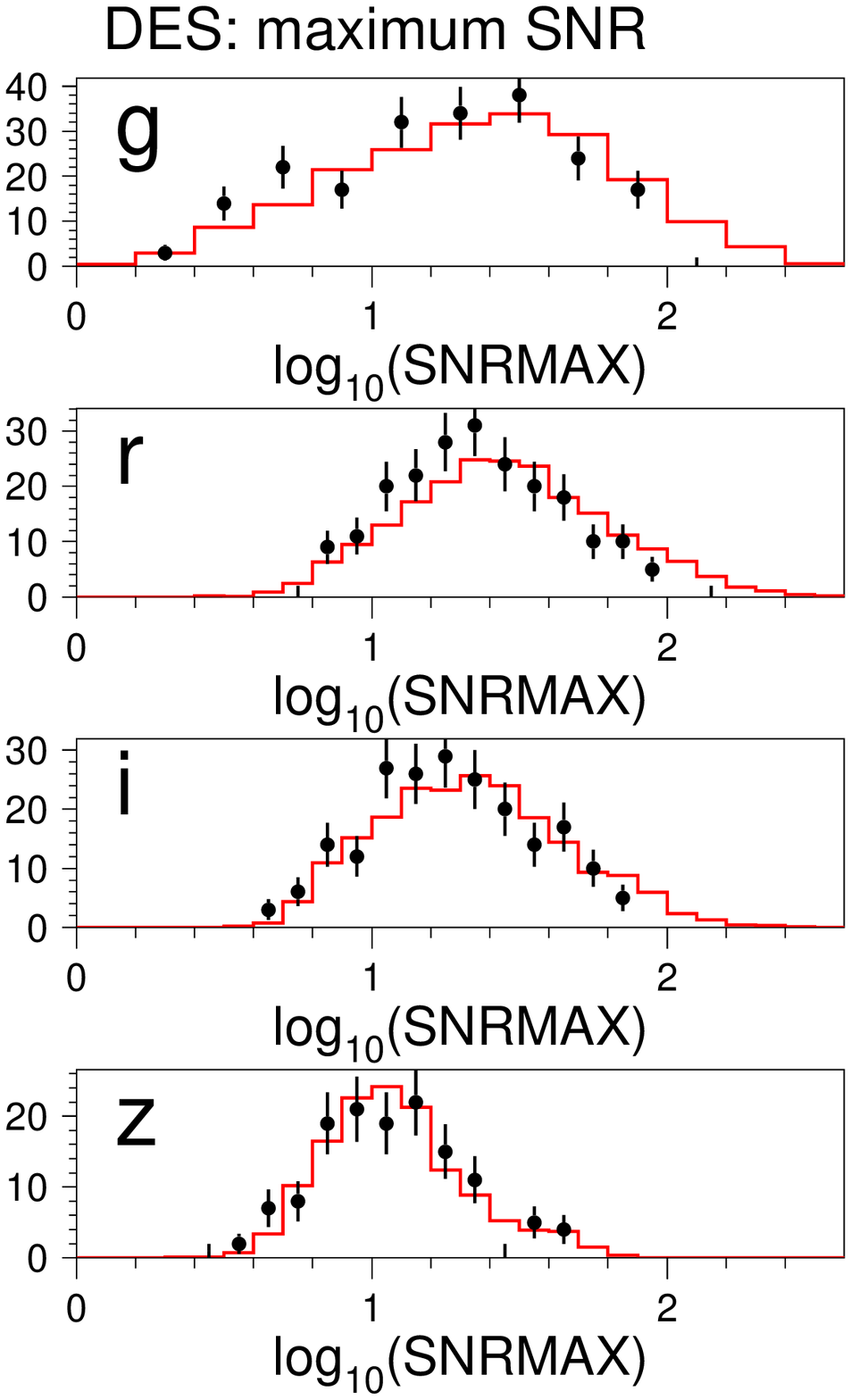}
\includegraphics[scale=0.38]{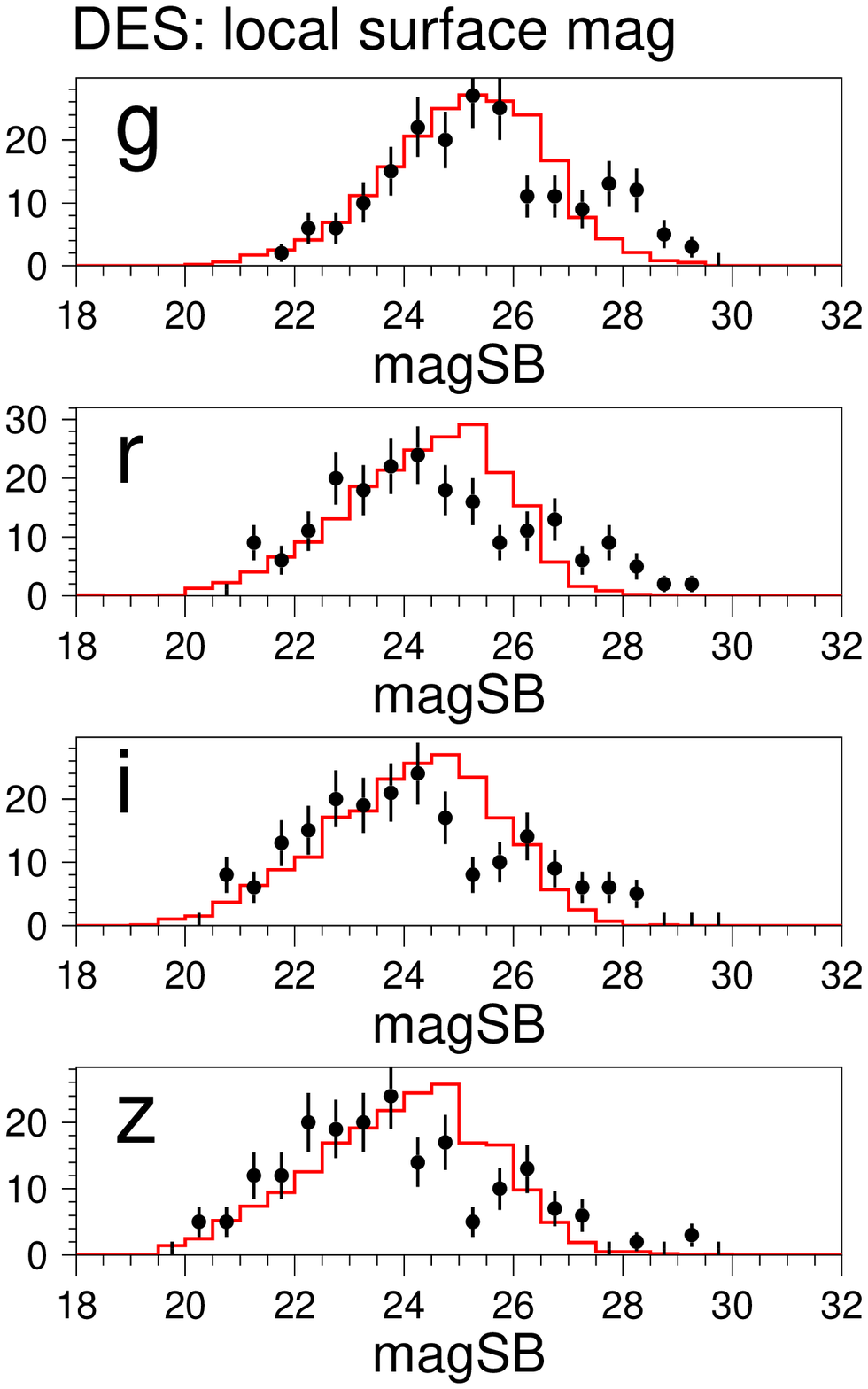}
   \end{center}
  \caption{
    Comparison of data (black dots) and simulation with G10 scatter model (red histogram) 
    for distributions in the \DESSN\ sample, where the simulation is scaled
    to have the same number of events as the data.
    Left column shows $\ipeak$, CMB redshift, Galactic extinction,  
    and maximum gap between \obss\ (rest frame).
    Middle column shows log of maximum SNR in each band. 
    Right column shows local surface mag in each band.
  }
\label{fig:ovdatasim_DES}  \end{figure*}

We apply light-curve fitting and selection requirements (cuts) that depend on 
\SALTII\ fitted parameters, SNR, and light curve sampling
(\S3.5 of \citealt{Brout2018_ANA}).
After applying these cuts,
data/simulation comparisons for \DESSN\ are shown in Fig.~\ref{fig:ovdatasim_DES}.
The $\ipeak$ distribution for data and simulation  are guaranteed to match because 
of the method for determining $\EFFspec$ in \S\ref{subsec:trigger_DESSN}.
The redshift agreement is not enforced, but is still excellent.
The next two distributions, $E(B-V)$ and maximum gap between \obss,
are also in excellent agreement, and this agreement validates the choice of 
random sky locations in the cadence library.  
The double peak structure of $E(B-V)$ is from the large sky separations
between groups of fields.

The middle column of Fig.~\ref{fig:ovdatasim_DES} compares the maximum SNR in each band, 
and these are the most difficult distributions to predict with the simulation. 
The comparisons look good, except for a slight excess in the simulation
for ${\rm SNR} > 100$.
The right column of Fig.~\ref{fig:ovdatasim_DES} compares the local surface brightness
mag in each band. There is good agreement in all bands for ${\rm SB}<24$,
For fainter hosts beyond the detection limit the agreement is much poorer, 
and is likely due to Malmquist bias for the limited co-add depth used in this analysis. 
Note that the poor agreement for faint hosts 
results in relatively small $\ERRSCALESIM$ errors because
$\ERRSCALESIM \to 1$ (Eq.~\ref{eq:sigF} and Fig.~\ref{fig:fluxerrscale}) 
as the underlying host becomes more faint,
and therefore the range of possible $\ERRSCALESIM$ corrections is smaller.

\begin{figure}
\centering
\begin{center}
   \includegraphics[scale=0.44]{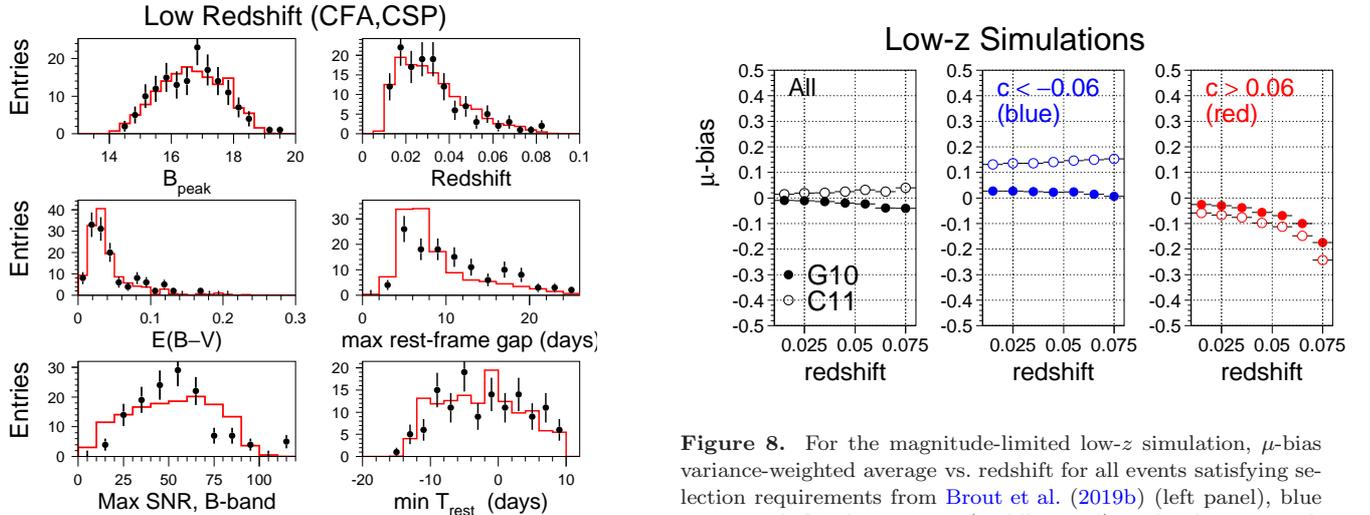}
   \end{center}
  \caption{
    Data/simulation comparisons for distributions in the \lowz\ sample,
    using the G10 intrinsic scatter model and magnitude-limited selection model.
  }
\label{fig:ovdatasim_LOWZ}  \end{figure}

Figure~\ref{fig:ovdatasim_LOWZ} shows data/simulation comparisons for the \lowz\ sample.
The $\Bpeak$ distributions are forced to match because of the method for determining $\EFFspec$.
The comparisons for redshift, $E(B-V)$ and minimum $\Trest$ show excellent agreement.
The comparisons for maximum gap between \obss\ (rest-frame) and maximum $B$-band SNR
indicate a slight discrepancy. The SNR agreement is poorer compared to \DESSN\
because we do not have the \obs\ information for the \lowz\ sample, and thus rely
on approximations  (\S\ref{sss:cadence_lowz})
to compute the noise in Eq.~\ref{eq:sigF}.

We have implemented \SALTII\ light curve fits on the simulations, 
and  Fig.~7  of \citet{Brout2018_ANA} shows data/simulation comparisons
for the \SALTII\ parameters ($m_B$, $x_1$, $c$) and their \uncs. 
The excellent agreement in these distributions adds confidence in our \mubias\ predictions.

\section{ Distance Modulus Bias vs. Redshift }
\label{sec:bias}

In one of our  \DESLOWZ\ cosmology analyses \citep{Brout2018_ANA}, 
we use the BBC method \citep{BBC} in which \mubias\ is characterized as a 
5-dimensional function of $\{z,x_1,c,\alpha,\beta\}$.
The first three parameters are observed, and $\{\alpha,\beta\}$ are
determined from the BBC fit.
Here we illustrate \mubias\ as a function 
of redshift for a variety of sub-samples, and also compare \mubias\ for the two 
intrinsic scatter models (G10,C11) from \S\ref{subsec:model_int}.
It is important to note that \mubias\ is not a correction for the SN magnitude,
but is a correction for fitted light curve parameters (describing the stretch, color and brightness) 
along with a correction for the impact of intrinsic scatter in which brighter events are preferentially 
selected in a magnitude-limited survey.

The true distance modulus is defined as $\mutrue$,
and the measured distance modulus ($\mu$) is determined  in the analysis
from \citet{Tripp1998},
\begin{equation}
    \mu =  -2.5\log(x_0) + \alpha x_1 - \beta c + \M~,
     \label{eq:mu}
\end{equation}
where $\{ x_0, x_1, c\}$ are fitted \SALTII\ light-curve parameters,
$\alpha$ and $\beta$ are the standardization parameters,
and $\M$  is an offset so that $\mu = \mutrue$ when the true values of 
$\{ x_0, x_1, c\}_{\rm true}$ are used in Eq.~\ref{eq:mu}. 
The distance modulus bias is defined as  {\mubias}~$\equiv \mu - \mutrue$.
The BBC method applies a {\mubias} correction  for each event,
and determines the following parameters in a fit to the entire sample:
$\alpha,\beta,\M$,  and a weighted-average bias-corrected distance modulus 
in discrete redshift bins.

We implement the BBC procedure on a simulated \DESLOWZ\ data sample
with $3{\times}10^4$ events  after applying the cuts from \S\ref{sec:ovdatasim}.
The \mubias\ thus has contributions from the \Diff\ trigger, \spec\ selection, and analysis cuts. 
We use a large `bias-correction'  sample with $\NSIMBIASCOR$ events after the same cuts.
Samples are generated with both the G10 and C11 intrinsic scatter model,
and the bias-correction sample with the correct intrinsic scatter model
is used on the data; the effect of using the incorrect model is discussed in 
\citet{Brout2018_ANA}.

To account for a \mubias\ dependence on $\alpha$ and $\beta$,
we generate the bias-correction sample on a $2\times 2$ grid of $\alpha\times \beta$ 
and use this grid for interpolation within the BBC fit.
The grid values are $\alpha = \{0.10,0.24 \}$,  
$\beta_{\rm G10} = \{2.7,3.5\}$, and $\beta_{\rm C11} = \{3.3,4.3\}$.

\begin{figure}
\centering
\begin{center}
   \includegraphics[scale=0.44]{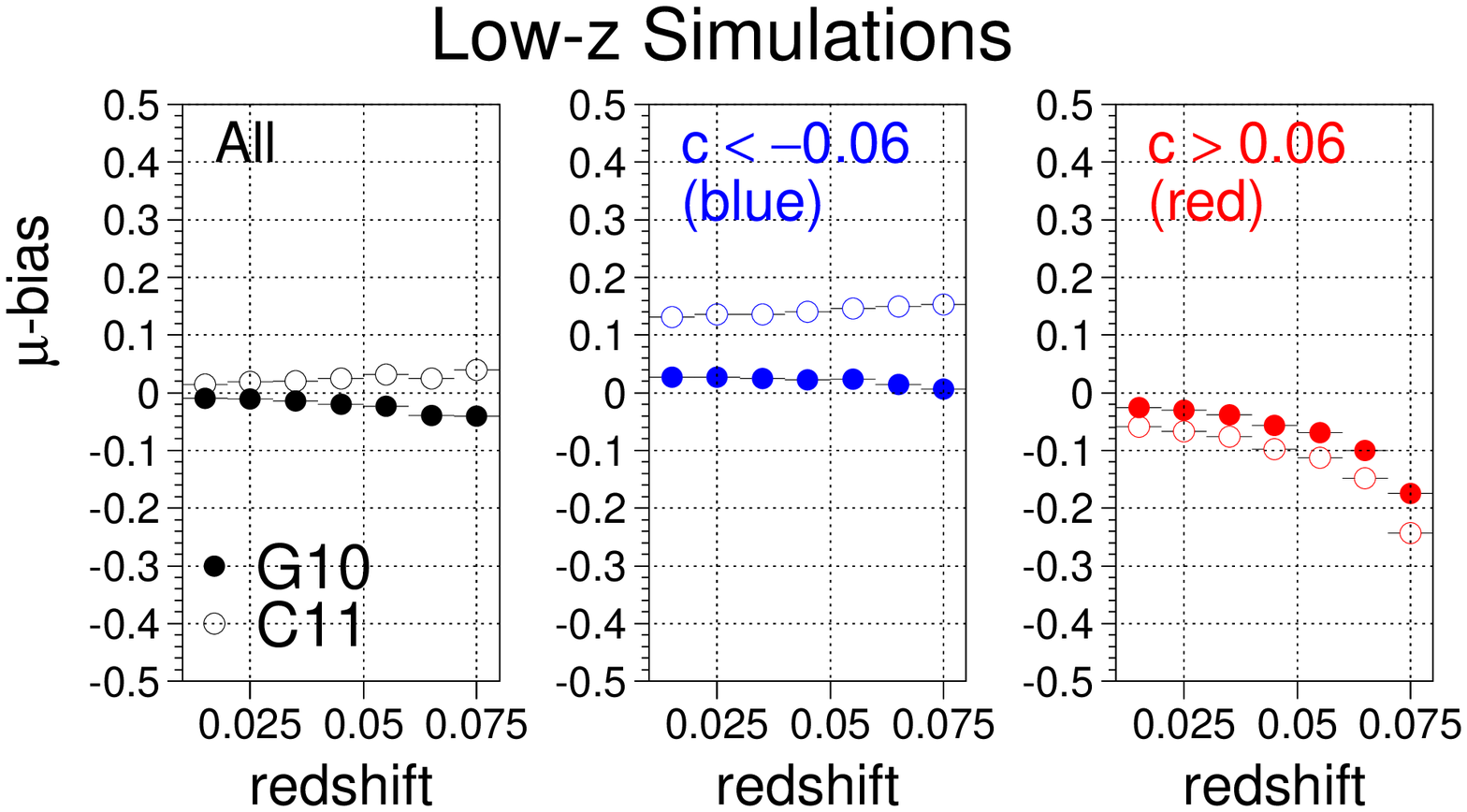}
   \end{center}
  \caption{
  For the magnitude-limited \lowz\ simulation, 
  $\mu$-bias variance-weighted average vs. redshift for all events satisfying selection requirements from \citet{Brout2018_ANA}
  (left panel), 
  blue events with fitted $c<-0.06$ (middle panel), and
  red events with fitted $c>0.06$ (right panel).
  Filled circles are with simulations using the G10 intrinsic scatter model;
  open circles are for the  C11 model.  
  }
\label{fig:mubias_lowz}  \end{figure}

\begin{figure}
\centering
\begin{center}
   \includegraphics[scale=0.44]{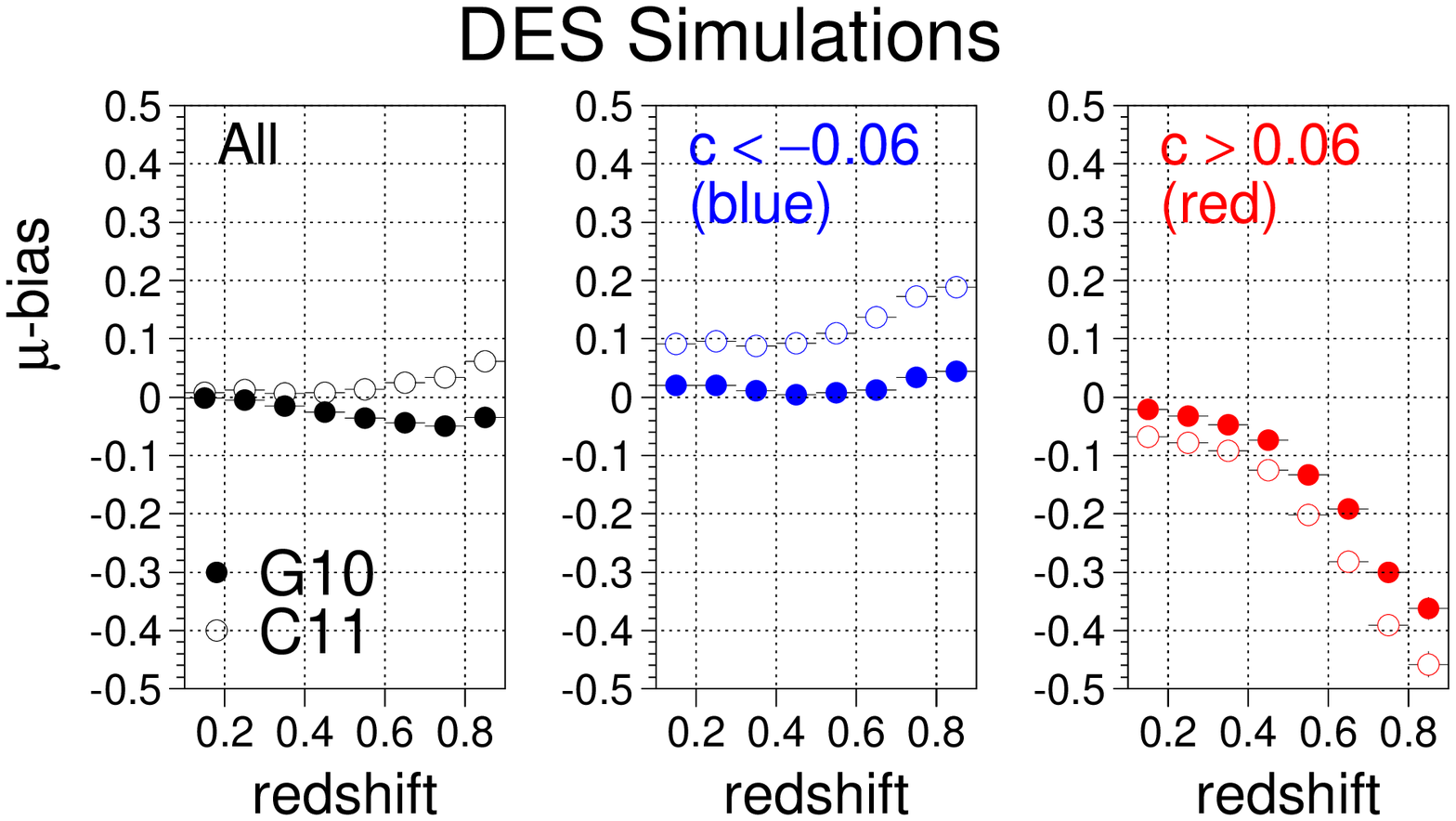}
   \end{center}
  \caption{
   Same as Fig.~\ref{fig:mubias_lowz}, but for \DESSN.
  }
\label{fig:mubias_des}  \end{figure}

The BBC-fitted values of $\alpha$ and $\beta$ are un-biased within their 5\% statistical \uncs,
and fitting with optional $z$-dependent slope parameters, 
$d\alpha/dz$ and $d\beta/dz$ are both consistent with zero.
$\M$ does not contribute to {\mubias} and therefore
the \mubias\ is caused by the fitted light curve parameters $\{ x_0, x_1, c\}$.
The $\mu$-bias versus redshift from the BBC fit 
is  shown in Figs.~\ref{fig:mubias_lowz}-\ref{fig:mubias_des}
for the \lowz\ and \DESSN\ samples, respectively.
The filled circles correspond to the G10 intrinsic scatter model,
and open circles correspond to C11.

The average \mubias\ (left panels) is zero at the lower end of the redshift range.  
At higher redshifts, \mubias\  depends on the intrinsic scatter model,
reaching $\sim \MAXMUBIASALL$~mag at the high-redshift range.
The middle and right panels of Figs.~\ref{fig:mubias_lowz}-\ref{fig:mubias_des}
show that  $\mu$-bias is much larger within restricted color ranges,
reaching up to $\MAXMUBIAS$~mag for the reddest ($c>0.06$) events.
All panels show a  \mubias\ difference between the G10 and  C11 models,
and this difference is largely due to the different parent color populations 
\citep{Scolnic2014color}: the C11 color population has a sharp cut-off on the blue side,
while the G10 population has a tail extending bluer than in the C11 model. 
These \mubias\ differences, along with differences in fitted $\alpha$ and $\beta$, 
are incorporated into the systematic \unc\ in \citet{Brout2018_ANA}. 

\newcommand{\RMSdc}{{\rm rms}(\Delta c)}
\newcommand{\RMSctrue}{{\rm rms}(c_{\rm true})}

The large \mubias\ for red events at higher redshift is because most of these
events are intrinsically blue, which are bright enough to be detected, 
but have poorly measured colors.  Intrinsically red events are fainter and thus tend
to be excluded at higher redshifts. 
To illustrate the size of the color \uncs\ for the \DESSN\ sample, 
we computed the rms on measured color minus true color, $\RMSdc$, 
and the rms of the true color population, $\RMSctrue$.
The ratio is $\RMSdc/\RMSctrue \sim 0.5$. 
Therefore the typical difference between measured and true color is
50\% of the size of the intrinsic color distribution. 
For redshifts $z>0.5$ this ratio increases to 0.7.
A similar exercise with the stretch parameter results in similar ratios.

\begin{figure}
\centering
\begin{center}
   \includegraphics[scale=0.44]{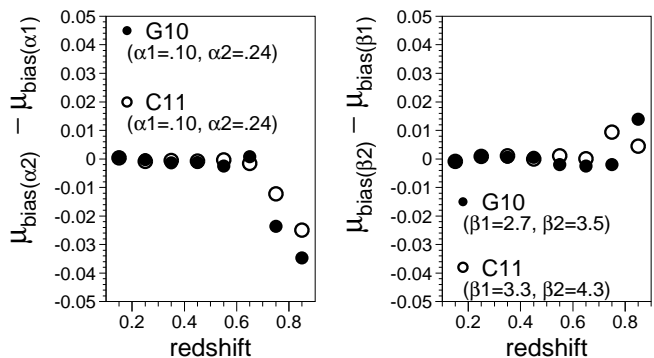}
   \end{center}
  \caption{
   For the \DESSN\ sample,
   left panel shows \mubias\ difference vs. redshift between $\alpha=0.10$ and $\alpha=0.24$.
   Right panel shows \mubias\ difference between different $\beta$ values:
   $\{2.7,3.3\}$ for G10 intrinsic scatter model (filled circles), 
   and $\{3.3,4.3\}$ for C11 (open circles).
  }
\label{fig:mubias_ab}  \end{figure}

As described in \S\ref{subsec:SALT2par}, a new feature in the BBC method is to 
account for the \mubias\ dependence on $\{\alpha,\beta\}$.
This dependence is shown in Fig.~\ref{fig:mubias_ab} for \DESSN.
Comparing simulations for $\alpha=0.10$ and $\alpha=0.24$ (nominal $\alpha\simeq 0.15$),
the \mubias\ difference reaches $0.03$~mag at high redshift, and is similar
for the two intrinsic scatter models (G10 and C11).
The right panel in Fig.~\ref{fig:mubias_ab} shows the \mubias\ difference with 
$\beta$ values differing by $\sim 1$; the maximum \mubias\ difference is 0.01~mag,
and is similar for both intrinsic scatter models.

\begin{figure}
\centering
\begin{center}
   \includegraphics[scale=0.44]{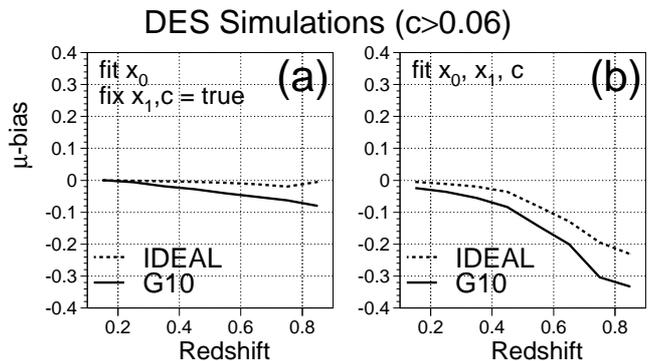}
   \end{center}
  \caption{
   For the simulated \DESSN\ sample,
   (a) shows \mubias\ vs. redshift for light curve fits that float $x_0$ only,
   while fixing stretch and color to their true values.
   (b) shows \mubias\ vs. redshift for nominal light curve fits.
   Dashed line is for the ideal simulation defined as having no intrinsic scatter;
   solid line uses G10 intrinsic scatter model.
  }
\label{fig:mubias_breakdown}  \end{figure}

We end this section by illustrating the contributions to  \mubias\ for
red events ($c>0.06$) in the right panel of Fig.~\ref{fig:mubias_des},
where \mubias\ reaches $\sim 0.4$~mag at the highest redshifts.
While high-redshift bias is often associated with Malmquist bias,
we show that \mubias\ is primarily associated with intrinsic scatter and light-curve fitting.
We begin with an ideal \DESSN\ simulation that has no intrinsic scatter,
and perform  light-curve fits in which only the amplitude $x_0$ is floated
while stretch and color ($x_1,c$) are assumed to be perfectly known.
Defining $m_0 = -2.5\log(x_0)$, \mubias\ and $m_0$-bias are the same.
The resulting \mubias\ is shown by the dashed curve in 
Fig.~\ref{fig:mubias_breakdown}a; 
this bias is only $\sim 0.01$~mag, a very small fraction of the \mubias\
in Fig.~\ref{fig:mubias_des}.
While there may be selection bias in the 2 detections contributing to the trigger (\S\ref{sec:trigger}),
the remaining few dozen epochs are not biased, and thus the majority
of \obss\ used to measure $x_0$ are un-biased.

The solid curve in Fig.~\ref{fig:mubias_breakdown}a  shows \mubias\
with the G10 intrinsic scatter model, and still fitting only for $x_0$.
In this case, \mubias\ increases considerably to  about 0.1~mag at the highest redshift,
and is a result of the strong brightness correlations among epochs and passbands.
While the true intrinsic scatter variations average to zero,
magnitude-limited \obss\ preferentially select positive brightness fluctuations,
which lead to non-zero \mubias.

Figure~\ref{fig:mubias_breakdown}b shows the same simulations,
but with  light curve fits that float all three parameters ($x_0,x_1,c$).
Compared with Fig.~\ref{fig:mubias_breakdown}a,
the \mubias\ is much larger, mainly because of the bias in fitted color.
Although this \mubias\ test is shown only for the the red events in Fig.~\ref{fig:mubias_des},
similar trends exist in all color ranges.

The statistical \uncs\ on these \mubias\ corrections are negligible.
Systematic \uncs\ in \citet{Brout2018_ANA}  are thus determined from changing input
assumptions such as the color and stretch populations, model of intrinsic scatter, 
and the value of the flux-\unc\ scale, $\ERRSCALESIM$.

\section{Conclusion}
\label{sec:fin}

The \SNANA\  simulation program has been under active development for a decade,
and has been used in several cosmology analyses to accurately simulate 
SN~Ia light curves and determine bias corrections for the distance moduli.
This work focuses on simulated bias corrections for the \DESLOWZ\ sample,
which combines  \specy\ confirmed SNe~Ia  from \DESSN\ and low-redshift samples.
Files used to make these corrections are available at {\urlDR}.

The \DESSN\ simulation includes three categories of detailed modeling:
(1) source model including the rest-frame SN~Ia SED,
cosmological dimming, weak lensing, peculiar velocity, and Galactic extinction,
(2) noise model accounting for  \obs\ properties (PSF, sky noise, zero point), 
host galaxy, and information derived from \NFAKE\ fake SN light curves overlaid 
on images and run through our image-processing pipelines,
(3) trigger model of single-visit detections, candidate logic, and \spec\ selection \eff.
The \lowz\ sample, however, does not include \obs\ properties,
and thus approximations are used to simulate this sample.
The quality of the simulation is illustrated by predicting observed
distributions (Figs.~\ref{fig:ovdatasim_DES}-\ref{fig:ovdatasim_LOWZ}),
and bias corrections on the distance moduli are shown in
Figs.~\ref{fig:mubias_lowz}-\ref{fig:mubias_des}.

The reliability of the bias corrections is only as good as the underlying
assumptions in the simulation. To properly propagate bias correction \uncs\ 
into systematic \uncs\ on cosmological parameters, \citet{Brout2018_SMP} evaluate
uncertainties for each of the 3 modeling categories above (source, noise, trigger). 
In addition to explicit assumptions such as those associated with the \SALTII\ model, 
one should always be aware of the implicit assumptions 
such as simulating SN properties (e.g., $\alpha$, $\beta$)
that are independent of redshift and host galaxy properties.

The simulations presented here are used to correct SN~Ia distance biases
in the \DESLOWZ\ sample \citep{Brout2018_ANA}, 
and these bias-corrected distances are  used to 
measure cosmological parameters  \citep{KEY2018}. 
These simulations also serve as a starting point for the analysis of the full DES 5-year 
photometrically classified sample, which will be significantly larger than the 
\DESLOWZ\ sample.

 \section{ Acknowledgements }
 \label{sec:Ack}
 
This work was supported in part by the Kavli Institute for Cosmological Physics at the University of Chicago 
through grant NSF PHY-1125897 and an endowment from the Kavli Foundation and its founder Fred Kavli.
This work was completed in part with resources provided by the 
University of Chicago Research Computing Center.
R.K. is supported by DOE grant DE-AC02-76CH03000.
D.S. is supported by NASA through Hubble Fellowship grant HST-HF2-51383.001 awarded by the
Space Telescope Science Institute, which is operated by the Association of
Universities for Research in Astronomy, Inc., for NASA, under contract NAS 5-26555.
The U.Penn group was supported by DOE grant DE-FOA-0001358 and NSF grant AST-1517742.
A.V.F.'s group at U.C. Berkeley is grateful for financial assistance from
NSF grant AST-1211916, the Christopher R. Redlich Fund, the TABASGO
Foundation, and the Miller Institute for Basic Research in Science.

Funding for the DES Projects has been provided by the U.S. Department of Energy, the U.S. National Science Foundation, the Ministry of Science and Education of Spain, 
the Science and Technology Facilities Council of the United Kingdom, the Higher Education Funding Council for England, the National Center for Supercomputing 
Applications at the University of Illinois at Urbana-Champaign, the Kavli Institute of Cosmological Physics at the University of Chicago, 
the Center for Cosmology and Astro-Particle Physics at the Ohio State University,
the Mitchell Institute for Fundamental Physics and Astronomy at Texas A\&M University, Financiadora de Estudos e Projetos, 
Funda{\c c}{\~a}o Carlos Chagas Filho de Amparo {\`a} Pesquisa do Estado do Rio de Janeiro, Conselho Nacional de Desenvolvimento Cient{\'i}fico e Tecnol{\'o}gico and 
the Minist{\'e}rio da Ci{\^e}ncia, Tecnologia e Inova{\c c}{\~a}o, the Deutsche Forschungsgemeinschaft and the Collaborating Institutions in the Dark Energy Survey. 

The Collaborating Institutions are Argonne National Laboratory, the University of California at Santa Cruz, the University of Cambridge, Centro de Investigaciones Energ{\'e}ticas, 
Medioambientales y Tecnol{\'o}gicas-Madrid, the University of Chicago, University College London, the DES-Brazil Consortium, the University of Edinburgh, 
the Eidgen{\"o}ssische Technische Hochschule (ETH) Z{\"u}rich, 
Fermi National Accelerator Laboratory, the University of Illinois at Urbana-Champaign, the Institut de Ci{\`e}ncies de l'Espai (IEEC/CSIC), 
the Institut de F{\'i}sica d'Altes Energies, Lawrence Berkeley National Laboratory, the Ludwig-Maximilians Universit{\"a}t M{\"u}nchen and the associated Excellence Cluster Universe, 
the University of Michigan, the National Optical Astronomy Observatory, the University of Nottingham, The Ohio State University, the University of Pennsylvania, the University of Portsmouth, 
SLAC National Accelerator Laboratory, Stanford University, the University of Sussex, Texas A\&M University, and the OzDES Membership Consortium.

Based in part on observations at Cerro Tololo Inter-American Observatory, National Optical Astronomy Observatory, which is operated by the Association of 
Universities for Research in Astronomy (AURA) under a cooperative agreement with the National Science Foundation.

The DES data management system is supported by the National Science Foundation under Grant Numbers AST-1138766 and AST-1536171.
The DES participants from Spanish institutions are partially supported by MINECO under grants AYA2015-71825, ESP2015-66861, FPA2015-68048, SEV-2016-0588, SEV-2016-0597, and MDM-2015-0509, 
some of which include ERDF funds from the European Union. IFAE is partially funded by the CERCA program of the Generalitat de Catalunya.
Research leading to these results has received funding from the European Research
Council under the European Union's Seventh Framework Program (FP7/2007-2013) including ERC grant agreements 240672, 291329, and 306478.
We  acknowledge support from the Australian Research Council Centre of Excellence for All-sky Astrophysics (CAASTRO), through project number CE110001020, and the Brazilian Instituto Nacional de Ci\^encia
e Tecnologia (INCT) e-Universe (CNPq grant 465376/2014-2).

This manuscript has been authored by Fermi Research Alliance, LLC under Contract No. DE-AC02-07CH11359 with the U.S. Department of Energy, Office of Science, Office of High Energy Physics. The United States Government retains and the publisher, by accepting the article for publication, acknowledges that the United States Government retains a non-exclusive, paid-up, irrevocable, world-wide license to publish or reproduce the published form of this manuscript, or allow others to do so, for United States Government purposes.


\bigskip\bigskip
\appendix
\section{ Additional Simulation Features for Future Analysis}
\label{sec:misc}

The focus of this work has been on simulating bias corrections and validation samples
for the \DESLOWZ\  SN~Ia cosmology analysis. 
Here we describe additional features of the \SNANA\ simulation that 
have been developed for future work, but are beyond the current scope of the \DESLOWZ\ 
analysis. This future work includes  extending the cosmology analysis to photometrically 
identified SNe~Ia,  more detailed systematics studies, determining the \eff\ for Bayesian 
cosmology fitting methods (e.g., \citealt{UNITY}), 
determining the \eff\ for SN rate studies,  and optimizing future surveys.
We end with a summary of missing features that would
be useful to add for future analysis work.

\subsection{SED Time Series}
\label{subsec:SED}

The \SALTII\ light curve model, which is designed for SN~Ia cosmology analyses,
is a rather complex semi-analytical model.
Most transient models, however, are much simpler. 
In addition to specialized SN~Ia 
models,\footnote{SN~Ia models in \SNANA\ include
      {\SALTII} \citep{Guy2010}, 
      {\bf\tt MLCS2k2} \citep{Jha2007}, and
      {\bf\tt SNOOPY} \citep{SNOOPY}. }
the \SNANA\ simulation works with arbitrary collections of SED time series.
Each event can be generated from a random SED time series,
or computed from parametric interpolation. For example, suppose
a set of $N_p$ parameters, $\vec{P} = \{ p_1,p_2, ... p_{N_p} \}$,
describes each SED time series. Each parameter ($p_i$) can be drawn
from a Gaussian distribution (or asymmetric Gaussian) and a full
covariance matrix to induce correlations. The SEDs on the parameter
grid are interpolated to the generated $\vec{P}$.

Examples include CC simulations to model contamination in photometrically identified 
SN~Ia samples \citep{K10_SNPCC,Rodney2012,BBC,Jones2017},
and simulating Kilonovae \citep{BK2013}
to model the search \eff\  \citep{Santos2016,Doctor2017},
and to predict discovery rates \citep{ScolnicKN17}.

An SED time series can also be useful for modeling SNe~Ia.
Examples include systematic studies on training the \SALTII\ model 
with simulated spectra \citep{Hsiao2007,Mosher2014}, 
and simulating spectra from SN~Ia explosion models \citep{Diemer2013,K13}.

\subsection{Light Curve Library for Galactic Transients }
\label{subsec:LCLIB}

Galactic transients can potentially contribute contamination in a photometrically identified 
SN~Ia sample. To model galactic transients, the simulation reads a  pre-computed 
`light curve library' of transient magnitudes versus time. 
The light curves can be recurring or non-recurring.
For recurring and long-lived non-recurring transients, 
the library specifies source magnitudes at epochs to use as templates for image-subtraction,
and the simulation accounts for source signal in the templates.
The subtracted fluxes can therefore be positive or negative.
To detect negative fluxes with ${\rm SNR}<0$, there is an option to 
define the detection efficiency as a function of $\vert {\rm SNR} \vert$. 
Each library light curve is overlaid on the survey time window, 
and overlapping  \obss\ in the cadence library are converted
into a measured flux and \unc. Readers are cautioned that this
model is relatively new, and has not yet been used in a publication.

\subsection{ Characterization of Detection Efficiency}
\label{subsec:MLscore}

For the \DESLOWZ\ analysis, the DES detection \eff\ was adequately characterized
as a function of SNR. In the next cosmology analysis with a much larger photometric sample,
we may need a more accurate description. 
In particular, we may need to characterize the \eff\ of a machine learning (ML)
requirement in \Diff\ that was used to reject image-subtraction artifacts \citep{autoScan}. 
The \SNANA\ data file structure includes a `{\tt PHOTPROB}'
entry for each epoch, which is intended to store information
such as an ML score. The simulation can generate ML scores
(between 0 and 1) based on an input probability map that depends
on SNR and/or $\mSB$. The input ML map should be generated
from fakes processed through the same pipeline as the data.
Since ML scores describe imaging data near the source, these scores 
are likely to be correlated among different epochs. 
A reduced correlation (0 to 1) can be provided
to introduce ML correlations.

While we have been characterizing anomalous effects as a function of $\mSB$,
we have begun exploring the dependence on $m-\mSB$, where $m$ is the source magnitude.
This source-to-galaxy flux ratio can be used to describe the
detection \eff\ or the ML map.

\subsection{ Characterization of Flux-Uncertainty Scale }
\label{subsec:ERRSCALE_more}

In  \S\ref{subsec:ERRSCALE} the flux-\unc\ scale, $\ERRSCALESIM$,
was defined as a function of 1 parameter: $\mSB$.
In future work we plan to investigate if $\ERRSCALESIM$
depends on other parameters.
The additional $\ERRSCALESIM$-dependent parameters in the simulation are: 
(1) SNR, 
(2) PSF, 
(3) MJD, 
(4) sky noise, 
(5) zero points, 
(6) galaxy magnitude, and 
(7) SN-host separation.
Additional parameters,  such as the source-to-galaxy flux ratio,
can be added with minor code modifications.

\subsection{Rate Models}

The following rate models can be used in the \SNANA\ simulation:
\begin{itemize}
   \item $R(z) = \alpha(1+z)^{\beta}$  with user-specified $\alpha,\beta$. Multiple $R(z)$ functions
         can be defined, each in a different redshift range.
   \item  $R(z) = A \cdot \int_{\infty}^z dz'{\SFR}(z') + B\cdot  {\SFR}(z)$          
            where {\SFR} is the star formation rate, $A$ is the amplitude of the delayed component,
            and $B$ is the amplitude of the prompt component \citep{ABrate2005,ABrate2006}.
   \item CC $R(z)$ measured with HST \citep{CC_S15}.
   \item Star formation $R(z)$ from \citet{MD14}, where user defines $R(0)$.
\end{itemize}

\subsection{ Redshift Dependent Input Parameters }
\label{subsec:zpar}

Since redshift evolution is a concern in cosmology analyses,
any simulation-input parameter can be given a redshift dependence:
$P \to P + p_1z + p_2z^2 + p_3z^3$, where $P$ is a user-specified 
simulation parameter and $p_{1,2,3}$ are user-defined parameters.
If a 3rd-order polynomial is not adequate, the simulation can read
an explicit $P(z)$ map in arbitrary redshift bins.

\subsection{ Population Parametrization }
\label{subsec:pop}

The \SALTII\ color and stretch populations are described by two 
asymmetric Gaussian profiles. The probability for color is defined as
\begin{eqnarray}
    P(c)  & \propto & \exp[ -(c-\cpeak)^2/ 2\sigplus^2 ]  ~~~ (c \ge \cpeak) \\
    P(c)  & \propto & \exp[ -(c-\cpeak)^2/ 2\sigminus^2 ]  ~~~ (c < \cpeak) 
\end{eqnarray}
and similarly for $P(x_1)$. A second asymmetric Gaussian can be added,
as described in Appendix~C of \citet{Pantheon} for the \lowz\ stretch distribution.

\subsection{ Inhomogeneous Distributions }
\label{subsec:not_isotropic}

The \DESLOWZ\ simulations assume an isotropic and homogeneous universe on 
all distance scales because of the random selection of sky coordinates in the
\obs\ library (\S\ref{subsec:cadence}) and the random generation of redshifts.
Large scale structure can be incorporated, but requires an external simulation 
to generate 3-dimensional (RA,DEC,$z$) galaxy locations. For each such galaxy,
the RA, DEC and redshift are used to create an entry in the \obs\ library.

Another application is to simulate transients corresponding to a posterior
from a gravitational wave (GW) event found by the 
Large Interferometer Gravity Wave Observatory \citep[LIGO]{LIGO_Singer2016,LIGO_Singer2016b}.
Drawing random events from the posterior described by RA, DEC and distance, 
each event corresponds to an entry in the \obs\ library.

\subsection{ Host Galaxy Library Features }
\label{app:hostlib}

A host-galaxy library (\HOSTLIB) was defined in \S\ref{subsec:host} to model 
additional Poisson noise and the local surface brightness. 
Additional \HOSTLIB\ features include:
\begin{itemize}
  \item mis-matched host redshift model for photometrically identified sample \citep{Jones2017},
  \item a weight map to assign SN magnitude offsets based on host-galaxy mass,
        or other properties such as specific star formation rate,
  \item photometric galaxy redshift (\ZPHOT) and Gaussian \unc\ (\ZPHOTERR),
            which must be computed externally from broadband filters,
  \item brightness distribution described with arbitrary sum of \Sersic\ profiles,
           each with its own index,
   \item correlation of host and SN properties by including \SALTII\ color and stretch 
             for each \HOSTLIB\ event.
\end{itemize}
            
\subsection{ Generating Spectra }
\label{subsec:spectra}

Ideally the modeling of \spec\ selection would include an analysis  of simulated spectra, 
but instead we empirically model this \eff\ as a function of peak magnitude.  
To begin the effort on modeling \spec\ selection,  the \SNANA\ simulation was enhanced 
to generate spectra for the WFIRST simulation study in \citet{Hounsell2018}.
Spectra are characterized by their SNR versus wavelength. 
They can be generated  at specific dates in the \obs\ library, or a random date 
can be selected in time windows with respect to peak brightness. 
This time-window can be specified in either the rest-frame or observer-frame, 
although the former is  more difficult to carry out in practice. 
Spectral slices can also be integrated and stored as broadband fluxes.              

Finally, a  high-SNR (\lowz) spectrum can be simulated at arbitrary redshift
to examine the expected SNR degradation versus distance.

\subsection{ Missing Features }
\label{subsec:missing}

We finish this section with a few features that are not included in the 
simulation, but might be useful in future analyses:

\begin{itemize}
  \item peculiar velocity covariances (currently all $\vpec$ are uncorrelated),
  \item galactic $E(B-V)$ covariance (currently all extinctions are uncorrelated),
  \item spectral PCA coefficients in the \HOSTLIB\ to model host 
           contamination in spectra,
  \item probability distribution for host-galaxy photometric redshifts 
       (instead of Gaussian-error approximation),
  \item  anomalous detection inefficiency from bright galaxies,
  \item  weak lensing magnification model (\S\ref{subsec:model_DL})
            accounting for correlations between events with small angular separations.
\end{itemize}

\section{Author Affiliations}
\label{sec:affil}

$^{1}$ Department of Astronomy and Astrophysics, University of Chicago, Chicago, IL 60637, USA\\
$^{2}$ Kavli Institute for Cosmological Physics, University of Chicago, Chicago, IL 60637, USA\\
$^{3}$ Department of Physics and Astronomy, University of Pennsylvania, Philadelphia, PA 19104, USA\\
$^{4}$ School of Mathematics and Physics, University of Queensland,  Brisbane, QLD 4072, Australia\\
$^{5}$ Lawrence Berkeley National Laboratory, 1 Cyclotron Road, Berkeley, CA 94720, USA\\
$^{6}$ The Research School of Astronomy and Astrophysics, Australian National University, ACT 2601, Australia\\
$^{7}$ Institute of Cosmology and Gravitation, University of Portsmouth, Portsmouth, PO1 3FX, UK\\
$^{8}$ ARC Centre of Excellence for All-sky Astrophysics (CAASTRO), Sydney, Australia\\
$^{9}$ School of Physics and Astronomy, University of Southampton,  Southampton, SO17 1BJ, UK\\
$^{10}$ University of Copenhagen, Dark Cosmology Centre, Juliane Maries Vej 30, 2100 Copenhagen O\\
$^{11}$ Korea Astronomy and Space Science Institute, Yuseong-gu, Daejeon, 305-348, Korea\\
$^{12}$ Harvard-Smithsonian Center for Astrophysics, 60 Garden St., Cambridge, MA 02138, USA\\
$^{13}$ INAF, Astrophysical Observatory of Turin, I-10025 Pino Torinese, Italy\\
$^{14}$ Millennium Institute of Astrophysics and Department of Physics and Astronomy, Universidad Cat\'{o}lica de Chile, Santiago, Chile\\
$^{15}$ South African Astronomical Observatory, P.O.Box 9, Observatory 7935, South Africa\\
$^{16}$ Space Telescope Science Institute, 3700 San Martin Drive, Baltimore, MD  21218, USA\\
$^{17}$ Department of Astronomy, University of California, Berkeley, CA 94720-3411, USA\\
$^{18}$ Miller Senior Fellow, Miller Institute for Basic Research in Science, University of California, Berkeley, CA  94720, USA\\
$^{19}$ Santa Cruz Institute for Particle Physics, Santa Cruz, CA 95064, USA\\
$^{20}$ Centre for Astrophysics \& Supercomputing, Swinburne University of Technology, Victoria 3122, Australia\\
$^{21}$ Department of Physics, University of Namibia, 340 Mandume Ndemufayo Avenue, Pionierspark, Windhoek, Namibia\\
$^{22}$ Harvard-Smithsonian Center for Astrophysics, 60 Garden St., Cambridge, MA 02138,USA\\
$^{23}$ Gordon and Betty Moore Foundation, 1661 Page Mill Road, Palo Alto, CA 94304,USA\\
$^{24}$ Sydney Institute for Astronomy, School of Physics, A28, The University of Sydney, NSW 2006, Australia\\
$^{25}$ Institute of Astronomy and Kavli Institute for Cosmology, Madingley Road, Cambridge, CB3 0HA, UK\\
$^{26}$ National Center for Supercomputing Applications, 1205 West Clark St., Urbana, IL 61801, USA\\
$^{27}$ Institute of Astronomy, University of Cambridge, Madingley Road, Cambridge CB3 0HA, UK\\
$^{28}$ Division of Theoretical Astronomy, National Astronomical Observatory of Japan, 2-21-1 Osawa, Mitaka, Tokyo 181-8588, Japan\\
$^{29}$ Institute of Astronomy and Astrophysics, Academia Sinica, Taipei 10617, Taiwan\\
$^{30}$ Observatories of the Carnegie Institution for Science, 813 Santa Barbara St., Pasadena, CA 91101, USA\\
$^{31}$ Cerro Tololo Inter-American Observatory, National Optical Astronomy Observatory, Casilla 603, La Serena, Chile\\
$^{32}$ Fermi National Accelerator Laboratory, P. O. Box 500, Batavia, IL 60510, USA\\
$^{33}$ Kavli Institute for Cosmology, University of Cambridge, Madingley Road, Cambridge CB3 0HA, UK\\
$^{34}$ LSST, 933 North Cherry Avenue, Tucson, AZ 85721, USA\\
$^{35}$ CNRS, UMR 7095, Institut d'Astrophysique de Paris, F-75014, Paris, France\\
$^{36}$ Sorbonne Universit\'es, UPMC Univ Paris 06, UMR 7095, Institut d'Astrophysique de Paris, F-75014, Paris, France\\
$^{37}$ Department of Physics \& Astronomy, University College London, Gower Street, London, WC1E 6BT, UK\\
$^{38}$ Kavli Institute for Particle Astrophysics \& Cosmology, P. O. Box 2450, Stanford University, Stanford, CA 94305, USA\\
$^{39}$ SLAC National Accelerator Laboratory, Menlo Park, CA 94025, USA\\
$^{40}$ Centro de Investigaciones Energ\'eticas, Medioambientales y Tecnol\'ogicas (CIEMAT), Madrid, Spain\\
$^{41}$ Laborat\'orio Interinstitucional de e-Astronomia - LIneA, Rua Gal. Jos\'e Cristino 77, Rio de Janeiro, RJ - 20921-400, Brazil\\
$^{42}$ Department of Astronomy, University of Illinois at Urbana-Champaign, 1002 W. Green Street, Urbana, IL 61801, USA\\
$^{43}$ Institut de F\'{\i}sica d'Altes Energies (IFAE), The Barcelona Institute of Science and Technology, Campus UAB, 08193 Bellaterra (Barcelona) Spain\\
$^{44}$ Institut d'Estudis Espacials de Catalunya (IEEC), 08034 Barcelona, Spain\\
$^{45}$ Institute of Space Sciences (ICE, CSIC),  Campus UAB, Carrer de Can Magrans, s/n,  08193 Barcelona, Spain\\
$^{46}$ Observat\'orio Nacional, Rua Gal. Jos\'e Cristino 77, Rio de Janeiro, RJ - 20921-400, Brazil\\
$^{47}$ Department of Physics, IIT Hyderabad, Kandi, Telangana 502285, India\\
$^{48}$ Department of Astronomy/Steward Observatory, 933 North Cherry Avenue, Tucson, AZ 85721-0065, USA\\
$^{49}$ Jet Propulsion Laboratory, California Institute of Technology, 4800 Oak Grove Dr., Pasadena, CA 91109, USA\\
$^{50}$ Instituto de Fisica Teorica UAM/CSIC, Universidad Autonoma de Madrid, 28049 Madrid, Spain\\
$^{51}$ Department of Astronomy, University of Michigan, Ann Arbor, MI 48109, USA\\
$^{52}$ Department of Physics, University of Michigan, Ann Arbor, MI 48109, USA\\
$^{53}$ Department of Physics, ETH Zurich, Wolfgang-Pauli-Strasse 16, CH-8093 Zurich, Switzerland\\
$^{54}$ Center for Cosmology and Astro-Particle Physics, The Ohio State University, Columbus, OH 43210, USA\\
$^{55}$ Department of Physics, The Ohio State University, Columbus, OH 43210, USA\\
$^{56}$ Harvard-Smithsonian Center for Astrophysics, Cambridge, MA 02138, USA\\
$^{57}$ Australian Astronomical Optics, Macquarie University, North Ryde, NSW 2113, Australia\\
$^{58}$ Departamento de F\'isica Matem\'atica, Instituto de F\'isica, Universidade de S\~ao Paulo, CP 66318, S\~ao Paulo, SP, 05314-970, Brazil\\
$^{59}$ George P. and Cynthia Woods Mitchell Institute for Fundamental Physics and Astronomy, and Department of Physics and Astronomy, Texas A\&M University, College Station, TX 77843,  USA\\
$^{60}$ Department of Astronomy, The Ohio State University, Columbus, OH 43210, USA\\
$^{61}$ Instituci\'o Catalana de Recerca i Estudis Avan\c{c}ats, E-08010 Barcelona, Spain\\
$^{62}$ Brandeis University, Physics Department, 415 South Street, Waltham MA 02453\\
$^{63}$ Instituto de F\'isica Gleb Wataghin, Universidade Estadual de Campinas, 13083-859, Campinas, SP, Brazil\\
$^{64}$ Computer Science and Mathematics Division, Oak Ridge National Laboratory, Oak Ridge, TN 37831\\


\bibliographystyle{mnras}
\bibliography{main}  


\bsp	
\label{lastpage}
\end{document}